\documentclass[letterpaper,11pt]{article}

\usepackage{amsmath,amssymb,amsthm,mathtools}
\usepackage{bbm}
\usepackage{enumitem}
\usepackage{url}
\usepackage[letterpaper,margin=1.25in,includefoot]{geometry}
\usepackage{titling}
\usepackage[hidelinks]{hyperref}
\usepackage{aliascnt}
\usepackage{csquotes}
\usepackage[backend=biber,style=apa]{biblatex}
\pretitle{\begin{center}\Large}
\postdate{\par\end{center}\vspace{-1em}}

\newtheorem{theorem}{Theorem}[section]

\newaliascnt{proposition}{theorem}
\newtheorem{proposition}[proposition]{Proposition}
\aliascntresetthe{proposition}

\newaliascnt{lemma}{theorem}
\newtheorem{lemma}[lemma]{Lemma}
\aliascntresetthe{lemma}

\newaliascnt{corollary}{theorem}
\newtheorem{corollary}[corollary]{Corollary}
\aliascntresetthe{corollary}

\theoremstyle{definition}
\newaliascnt{definition}{theorem}
\newtheorem{definition}[definition]{Definition}
\aliascntresetthe{definition}

\newaliascnt{example}{theorem}
\newtheorem{example}[example]{Example}
\aliascntresetthe{example}

\theoremstyle{plain}
\newtheorem{claim}{Claim}[section]

\numberwithin{equation}{section}
\newcommand{\one}{\mathbbm 1}
\newcommand{\Rev}{\mathrm{Rev}}
\newcommand{\SREV}{\mathrm{SREV}}

\title{Optimality of Separate Selling with Finite-Support Values
\thanks{I thank Masato Eguchi, Tohya Sugano, Keita Kuwahara, Meina Takahashi,
Kenji Utagawa, Michihiro Kandori, Hiroki Shinozaki, and Yuichi Yamamoto for helpful comments.}
}
\author{Sho Mikami\thanks{Graduate School of Economics, The University of Tokyo.
Email: \href{mailto:mikami-sho910@g.ecc.u-tokyo.ac.jp}{\nolinkurl{mikami-sho910@g.ecc.u-tokyo.ac.jp}}.}}
\date{November 4, 2026}

\begin{document}
\maketitle

\begin{abstract}
We characterize when separate selling is exactly revenue-optimal in
multi-item auctions with additive buyers and independent, finitely
supported valuations. For each item, the valuation distribution is
identical across buyers, but it may differ across items. We show that,
provided at least two items have a strictly positive non-top support
value, separate selling is optimal for any number of buyers ($n\ge2$)
if and only if \emph{top selling}, which sells each item
only at its highest possible value, is optimal for a single buyer. This
equivalence holds under both dominant-strategy incentive compatibility
and Bayesian incentive compatibility. We then characterize the
optimality of top selling in the one-buyer problem through the existence
of a particular class of dual flows and derive a necessary
condition and several sufficient conditions on the value distributions
for its optimality. These conditions reveal a common economic feature,
which we call \emph{top-value isolation}: high non-top valuations must be
sufficiently far below the highest value or carry sufficiently little
probability mass relative to the probability mass at the highest value.


\end{abstract}

\section{Introduction}
\label{sec:introduction}

Consider a seller who owns several heterogeneous items and faces multiple
buyers. Each buyer has a valuation for each item, and values a
bundle of items by the sum of these valuations. The seller does not observe
the buyers' valuations but knows their probability distributions. We assume
that valuations are independent across buyers and items. For each item, the
valuation distribution is identical across buyers, but it may differ across
items.

How should the seller sell these items to maximize expected revenue? In
particular, under what conditions is \emph{separate selling}, which runs an
optimal auction for each item independently, an optimal way to sell the items?

Revenue maximization with multiple buyers and multiple items is known to be
one of the most difficult problems in mechanism design. An explicit
characterization of optimal mechanisms is not available for general
distributions, and the problem has been solved analytically only in a limited
number of cases, including special environments in which valuations take only
two possible values. There are two main sources of this difficulty.

The first difficulty is that multiple items allow more detailed price discrimination. Even with one buyer, a mechanism can use values for other items to determine whether an item is allocated and how much the buyer pays in total. Offering individual items and discounted bundles can give buyers with the same value for one item different terms for obtaining it, depending on their values for other items. Lotteries add further flexibility by varying the probabilities of receiving items. \textcite{HartReny2015} show that randomization can yield strictly higher revenue than any deterministic mechanism. Tailoring offers to combinations of item values can therefore raise revenue above separate selling and makes the optimal multi-item mechanism difficult to characterize even for one buyer.

The second difficulty arises from competition among buyers. In the single-item problem, \textcite{Myerson1981} shows that the seller must account for the revenue forgone by allocating the item to one buyer rather than another. This opportunity cost can be captured by ironed virtual valuations: the optimal auction compares these valuations across buyers and with the option of not selling. With multiple items, however, it is much harder to summarize the effects of other buyers' reports through similarly simple opportunity costs and to ensure that these costs are consistent with allocation feasibility throughout the mechanism.

In the multi-buyer, multi-item setting, these two difficulties interact,
making the characterization of optimal mechanisms even more challenging. Given
these difficulties, a natural approach is to study when simple selling
mechanisms are optimal, rather than trying to characterize the optimal
mechanism directly. In particular, for the single-buyer, multi-item problem,
two simple mechanisms have received considerable attention: \emph{separate
selling}, which sells each item independently, and \emph{grand bundling},
which sells all items together as a single bundle
\parencite[see, e.g.,][]{HartNisan2017,BabaioffImmorlicaLucierWeinberg2020}.

However, a negative result is known regarding the exact optimality of separate
selling. \textcite{McAfeeMcMillanWhinston1989} study the problem of selling
two independent items to a single buyer. They show that, under certain
continuity conditions on the distributions, \emph{mixed bundling} generates
strictly higher revenue than separate selling. Mixed bundling is a selling
method that offers the buyer the option to purchase each item separately, as
well as the option to purchase multiple items together at a discounted price.

Does the non-optimality of separate selling continue to hold when multiple
buyers compete for the items? One might expect competition among buyers to
make separate selling more likely to be optimal. Indeed,
\textcite{FeldmanFriedlerRubinstein2018} show that separate selling becomes
asymptotically optimal as the number of buyers grows relative to the number
of items. In the finite-support setting considered here, as the number of
buyers increases, it becomes increasingly likely that at least one buyer has the highest
possible valuation for each item. Separate selling then generates revenue close to the maximum
feasible surplus, leaving little scope for additional gains from selling the items jointly.

In fact, separate selling can be exactly optimal in finite-support environments,
even with multiple buyers. For two items with i.i.d. two-point values,
\textcite{Yao2017} shows that separate selling is optimal when the higher
value is sufficiently far above the lower value. We investigate
how far this possibility extends by characterizing the independent
finite-support item distributions for which separate selling is optimal.

Our first result characterizes the optimality of separate selling with multiple buyers through a single-buyer problem. Suppose that at least two items have a strictly positive non-top support value. Then, for any number of buyers \(n\ge 2\), separate selling is optimal if and only if \emph{top selling}, which sells each item only at its highest possible value, is optimal for a single buyer. This equivalence holds under both dominant-strategy incentive compatibility (DSIC) with ex-post individual rationality and Bayesian incentive compatibility (BIC) with interim individual rationality. Thus, the distributions for which separate selling is exactly optimal do not depend on the number of buyers once there are at least two.

The theorem implies that separate selling can be optimal with multiple buyers only if selling each item exclusively at its highest possible value is optimal. In this sense, the theorem provides a negative result, imposing a strong restriction on the distributions for which separate selling can be optimal. Moreover, although separate selling becomes asymptotically optimal as the number of buyers grows, the class of distributions for which it is exactly optimal does not expand once there are at least two buyers.

Why can selling below the highest possible value make separate selling
suboptimal? Such sales can leave information rents to buyers with higher
valuations, who might otherwise benefit from reporting lower values. With
multiple buyers and items, however, the seller can use reports about one item
to change the allocation of another. For example, when two buyers are tied
for one item, the seller can use their reports about another item to decide
who receives it. This creates room to modify allocations and payments while
preserving incentive compatibility. The seller can then increase revenue
either by reducing information rents or by selling items that would otherwise
remain unsold without increasing expected rents. Under top selling, by
contrast, each item is sold only at its highest possible value and at that
price, leaving no information rents to exploit.

Our non-optimality argument differs from the single-buyer
mixed-bundling argument of
\textcite{McAfeeMcMillanWhinston1989}.
Their revenue improvement comes from offering a discounted
bundle alongside separately priced items.
In contrast, our proof does not rely on bundling:
with multiple buyers, allowing the allocation of one item
to depend on reports about another creates opportunities
to reduce information rents or introduce profitable
lottery sales while preserving incentive compatibility.

We next examine how restrictive this requirement is by characterizing
top-selling optimality for general finite-support distributions. We first
provide an exact characterization in terms of feasible \emph{top-preserving
flows}, and then derive conditions directly on the marginal distributions.
These conditions reveal that separate selling can be optimal only when high
non-top valuations are sufficiently far below the highest possible value or
sufficiently rare relative to it. We refer to this property as
\emph{top-value isolation}.

For example, under Myerson regularity, we obtain a simple necessary and sufficient condition for separate-selling optimality. Let $K_j$ denote the number of support values of item $j$.
Let $v_{j,K_j-1}$ and $\bar v_j$ denote its second-highest
and highest values, respectively, and let $p_j$ denote
its probability mass function.
Then, under our standing support assumption, separate selling is optimal with two or more buyers if and only if
\[
\sum_{j\in[m]}
\frac{
p_j(v_{j,K_j-1})v_{j,K_j-1}
}{
p_j(\bar v_j)(\bar v_j-v_{j,K_j-1})
}
\leq 1.
\]
This inequality makes the restriction explicit: for each item, the probability mass at the second-highest value must be sufficiently small relative to that at the top value, unless the gap between the two values is sufficiently large. Moreover, these restrictions accumulate across items, making optimality harder to achieve as more items are added.

\paragraph{Related literature.}
\textcite{Yao2017} derives exact optimal revenues under DSIC and BIC for two
items and any $n\ge2$ when all buyer--item values are i.i.d.\ draws from a
common two-point distribution. In particular, the results show that separate
selling can be exactly optimal with multiple buyers. Subsequent studies
extend exact optimal mechanism design to other binary environments.
\textcite{MertzanidisTerzoglou2026} obtain optimal BIC mechanisms for multiple
items or for two-item settings with buyer- or item-specific distributions.
\textcite{LiuLiuWang2026} characterize an optimal deterministic DSIC mechanism
among randomized mechanisms for i.i.d.\ two-point values and arbitrary
numbers of buyers and items. They also establish computational hardness
results for general finite-support distributions. Rather than characterizing
optimal mechanisms in these environments, we characterize the distributions
for which separate selling is optimal, allowing arbitrarily many support
values and item-specific distributions. Under Myerson regularity, our
characterization recovers the optimality region identified by
\textcite{Yao2017}.

\textcite[Corollary~5.1]{Zuo2017} shows that, when
all buyer--item values are i.i.d.\ and zero lies in
their common support, equality of optimal DSIC and
BIC revenues is equivalent to separate-selling
optimality for $n\ge3$.
Under our standing support assumption, our theorem
shows that separate selling is optimal under DSIC
and ex-post IR if and only if it is optimal under
BIC and interim IR.
This equivalence holds for every $n\ge2$ and allows
item-specific finite-support distributions.

For multi-bidder environments with continuously distributed values that are
symmetric across buyers and independent across items,
\textcite{JehielMeyerterVehnMoldovanu2007} construct DSIC mixed bundling
auctions that strictly outperform selling each item through an efficient
auction and, under additional conditions, selling each item through a
second-price auction with an optimal reserve price.
\textcite{KolesnikovSandomirskiyTsyvinskiZimin2022} provide an alternative
proof, using duality, that separate selling is suboptimal among BIC and
interim individually rational mechanisms when item values are independent
and have strictly positive, continuously differentiable densities.

Related studies examine the optimality of separate selling under different
assumptions. \textcite{Carroll2017} shows that separate screening is optimal
in a robust setting where the seller knows the marginal distributions but
evaluates mechanisms against the worst-case joint distribution consistent
with those marginals. In contrast, we assume independence across items and
characterize when separate selling is optimal for the resulting joint
distribution. \textcite{Kwak2026} studies a single buyer whose valuations for
all items increase with the buyer's type and characterizes when separate
monopoly pricing is optimal for every distribution of that type. Our analysis
instead focuses on given independent finite-support distributions.

A related line of research evaluates simple selling mechanisms through
approximation guarantees rather than exact optimality. In the single-buyer
setting with independent item values,
\textcite{CaiGuptaJiangLiawMehtaVelegkasWang2026},
\textcite{GiannakopoulosHahn2026Determinism}, and \textcite{Huang2026}
strengthen revenue guarantees for separate selling and grand bundling.
\textcite{Huang2026} also establishes competition guarantees for grand-bundle
auctions under regularity. These studies quantify the revenue loss from
restricting attention to simple mechanisms; in contrast, we identify the
distributions for which separate selling incurs no loss, including with
multiple buyers.

Our analysis is also related to the duality literature on multidimensional
mechanism design. \textcite{DaskalakisDeckelbaumTzamos2017} develop a
strong-duality framework for the multiple-good monopoly problem,
characterizing optimality through stochastic dominance and optimal
transport. \textcite{CaiDevanurWeinberg2021} interpret dual variables
associated with incentive constraints as flows between types and use the
induced virtual values to bound revenue.

More recently, \textcite{JiangParkesWang2026} construct computational dual revenue
certificates for multi-item, multi-bidder auctions under DSIC. Drawing on
the flow interpretation of \textcite{CaiDevanurWeinberg2021}, we instead
characterize one-buyer top-selling optimality exactly by the existence of
a feasible \emph{top-preserving flow}. We then derive conditions directly
on the marginal distributions that reveal how restrictive
separate-selling optimality is.

Section~\ref{sec:environment} presents the model.
Section~\ref{sec:reductions} establishes the one-buyer characterization
under DSIC and BIC, together with the flow criterion.
Section~\ref{sec:optimality} derives distributional conditions.
Section~\ref{sec:conclusion} concludes; Appendix~A contains supporting
results and proofs.

\section{Model}
\label{sec:environment}

For a positive integer $k$, write $[k]:=\{1,\ldots,k\}$ and
$[0]:=\varnothing$. There are $m\ge2$ heterogeneous items and
$n\ge1$ risk-neutral buyers. Buyer $i\in[n]$ values item $j\in[m]$
at $v_{ij}$. Values are independent across all buyer-item pairs.
For each item $j$, the values $v_{ij}$ have the same marginal
distribution $F_j$ across buyers; $F_j$ may differ across items.
Its support is
\[
V_j=\{v_{j,1}<\cdots<v_{j,K_j}\},
\]
where $K_j\ge2$, $v_{j,1}\ge0$, and $v_{j,K_j}=\bar v_j$.
The probability mass function $p_j$ satisfies
$\sum_{k\in[K_j]}p_j(v_{j,k})=1$ and
$p_j(v_{j,k})>0$ for every $k\in[K_j]$.

Our standing assumption is that at least two items have a strictly
positive non-top support value.
Equivalently, $v_{j,K_j-1}>0$ for at least two $j\in[m]$.
\footnote{If this assumption fails,
all but at most one item have support $\{0,\bar v_j\}$, and separate selling
is optimal under DSIC and ex-post individual rationality for every $n\ge1$; see
Lemma~\ref{lem:degenerate-items-separate}. Consequently, without the standing
assumption, separate selling can be optimal even when top selling is suboptimal
for the exceptional item.}
Auxiliary problems involving one item or a subset of items need not
satisfy this assumption.

Write $\mathbf F=(F_1,\ldots,F_m)$ and $\mathcal V:=\prod_{j=1}^m V_j$.
A buyer's type belongs to $\mathcal V$, and the type profile $v=(v_i)_{i \in [n]}$ belongs
to $\mathcal V^n$. For a buyer's type $t \in \mathcal V$, write
$\mathrm{Pr}(t):=\prod_{j=1}^m p_j(t_j)$.
For a type profile $v\in\mathcal V^n$, write
$\mathrm{Pr}(v):=\prod_{i\in[n]}\mathrm{Pr}(v_i)$.
The lowest and highest types are
$\underline v:=(v_{1,1},\ldots,v_{m,1})$ and
$\bar v:=(\bar v_1,\ldots,\bar v_m)$, respectively.

A direct mechanism $(x,\tau)$ consists of allocations
$x:\mathcal V^n\to[0,1]^{nm}$ and nonnegative payments 
$\tau:\mathcal V^n\to\mathbb R_+^n$.
Here $x_{ij}(v)$ is the probability that buyer $i \in [n]$ receives item $j \in [m]$.
Allocations and payments, and hence the DSIC and IR conditions below,
are evaluated in expectation over the mechanism's randomization,
conditional on the reported profile.
Write $x_i(v):=(x_{ij}(v))_{j=1}^m$ for buyer $i$'s allocation vector.
Buyers have additive values and quasi-linear truthful utility
$u_i(v):=\sum_{j\in[m]}v_{ij}x_{ij}(v)-\tau_i(v)$ for $v \in \mathcal V^n$.
\emph{Feasibility} requires
\[
\sum_{i\in[n]} x_{ij}(v)\le1
\]
for every $j\in[m]$ and $v\in\mathcal V^n$.
\emph{Dominant-strategy incentive compatibility}, abbreviated DSIC, requires
\[
u_i(t,v_{-i})
\ge
u_i(\hat t,v_{-i})+(t-\hat t)\cdot x_i(\hat t,v_{-i})
\]
for every $i\in[n]$, $t,\hat t\in\mathcal V$, and
$v_{-i}\in\mathcal V^{n-1}$. 
For one-buyer problems, we refer to the corresponding incentive-compatibility condition simply as IC.
\emph{Ex-post individual rationality}, abbreviated IR, requires $u_i(v)\ge0$
for every $i\in[n]$ and $v\in\mathcal V^n$.
Let $\mathcal M_{n,m}$ be the set of feasible DSIC and IR mechanisms.
The seller's optimal revenue is
\[
\Rev_n(\mathbf F):=
\max_{(x,\tau)\in\mathcal M_{n,m}}
\mathbb E_v\left[\sum_{i\in[n]}\tau_i(v)\right].
\]
The feasible set is closed and bounded: allocations lie in $[0,1]$,
and IR and nonnegative payments imply
$0\le\tau_i(v)\le\sum_{j\in[m]}\bar v_j$ at every profile.
Thus the maximum is attained. A mechanism attaining it is called
\emph{optimal}.

\emph{Bayesian incentive compatibility}, abbreviated BIC, requires
\[
\mathbb E_{v_{-i}}\left[u_i(t,v_{-i})\right]
\ge
\mathbb E_{v_{-i}}\left[
u_i(\hat t,v_{-i})+(t-\hat t)\cdot x_i(\hat t,v_{-i})
\right]
\]
for every $i\in[n]$ and $t,\hat t\in\mathcal V$.
\emph{Interim individual rationality}, abbreviated BIR, requires
\[
\mathbb E_{v_{-i}}\left[u_i(t,v_{-i})\right]\ge0
\]
for every $i\in[n]$ and $t\in\mathcal V$.
The expectations are over the other buyers' values, whose distribution
does not depend on buyer $i$'s value. Let $\mathcal M^B_{n,m}$ be the set
of feasible BIC and BIR mechanisms with nonnegative payments. Its optimal
revenue is
\[
\Rev^B_n(\mathbf F):=
\max_{(x,\tau)\in\mathcal M^B_{n,m}}
\mathbb E_v\left[\sum_{i\in[n]}\tau_i(v)\right].
\]
This maximum is attained: BIR bounds each interim expected payment by
the interim allocated value, and every opponents' profile has positive
probability, so nonnegative payments are bounded at every profile.
The finite constraint set is closed.

Nonnegative payments are without loss for optimal revenue under either
set of conditions. Under DSIC, the one-buyer normalization in the proof
of Lemma~\ref{lem:one-buyer-dual} applies at each fixed opponents'
profile. Under BIC and BIR, each buyer's interim allocations and
payments form an IC and IR menu. Normalize that menu and charge its
resulting nonnegative payment at every opponents' profile.

For a single-item distribution $F_j$, let
$\mathcal M_{n,1}$ denote the corresponding set of
feasible DSIC and IR mechanisms, and let
$\Rev_n(F_j)$ denote the optimal revenue.

\emph{The separate-selling mechanism} runs an optimal $n$-buyer
single-item auction for each marginal $F_j$. It is DSIC and IR, and hence
BIC and BIR.\footnote{The
DSIC property of separate selling relies on additivity. For example,
separate selling need not be DSIC for unit-demand buyers.} We denote the
expected revenue of the separate-selling mechanism by
\[
\SREV_n(\mathbf F):=\sum_{j\in[m]}\Rev_n(F_j).
\]
Our objective is to determine when separate selling is optimal under
either set of incentive and participation conditions.

\section{One-buyer reduction and flow characterization}
\label{sec:reductions}

\subsection{Separate selling and one-buyer top selling}

\begin{definition}[Top selling]
\label{def:top-selling}
For each item $j\in[m]$, the single-item top-selling mechanism
$M_j^{\mathrm{top}}\in\mathcal M_{n,1}$ operates as follows.
If at least one buyer reports $\bar v_j$, exactly one such buyer is
selected by a fixed tie-breaking rule, receives the item, and pays
$\bar v_j$. All others pay zero for that item.
If no buyer reports $\bar v_j$, the item is not sold and all buyers pay
zero for that item.
The top-selling mechanism $M^{\mathrm{top}}\in\mathcal M_{n,m}$ runs
these single-item mechanisms separately for all items.
\end{definition}
For each item $j \in [m]$, top selling yields revenue $\bar v_j$ whenever 
at least one buyer reports the highest value $\bar v_j$, an event that occurs 
with probability $1-(1-p_j(\bar v_j))^n$. 
Hence its expected revenue is
\[
\sum_{j\in[m]} \bar v_j\left[1-(1-p_j(\bar v_j))^n\right].
\]
For each item, top selling is a feasible single-item mechanism, 
so its revenue is no greater than the optimal single-item revenue. 
Summing across items gives the first inequality below.
Separate selling is feasible under DSIC and IR, and every DSIC and IR
mechanism is BIC and BIR. Thus,
\[
\sum_{j\in[m]} \bar v_j\left[1-(1-p_j(\bar v_j))^n\right]
\le \SREV_n(\mathbf F)
\le \Rev_n(\mathbf F)
\le \Rev^B_n(\mathbf F).
\]

\begin{theorem}[One-buyer reduction under DSIC and BIC]
\label{thm:separate-top-single-buyer}
Suppose at least two items have a strictly positive non-top support value,
and let $n\ge2$.
The following statements are equivalent.
\begin{enumerate}[label=(\roman*)]
\item Separate selling is optimal under BIC and BIR:
$\SREV_n(\mathbf F)=\Rev^B_n(\mathbf F)$.
\item Separate selling is optimal under DSIC and IR:
$\SREV_n(\mathbf F)=\Rev_n(\mathbf F)$.
\item Top selling is optimal for $n$ buyers under DSIC and IR.
\item Top selling is optimal for one buyer.
\end{enumerate}
\end{theorem}

The implication (i)$\Rightarrow$(ii) follows from
$\SREV_n(\mathbf F)\le\Rev_n(\mathbf F)\le\Rev^B_n(\mathbf F)$.
The central implication (ii)$\Rightarrow$(iii) shows that optimality of
separate selling under DSIC and IR forces top selling to be optimal for
$n$ buyers. Under separate
selling, directly reducing a buyer's rent at a high type may violate
a binding DSIC constraint. Yet the mechanism allocates each item without
using reports for other items. If an optimal single-item auction
allocates an item below its highest value, changing its allocation
among tied buyers according to another item's reports creates the
DSIC slack needed to reduce rent. When two items have such sales, the
expected sum of buyers' values times their allocation changes is zero
by buyer symmetry and independence across items, so the rent reduction
raises revenue. When only one item has such sales, a lottery for the
other item raises expected payments.
The proof verifies feasibility and every binding DSIC constraint directly;
finiteness allows a common small perturbation that preserves the
remaining strict DSIC constraints, IR, and nonnegative payments.

For (iii)$\Rightarrow$(iv), a one-buyer improvement over top selling
can be embedded when all other buyers report below the highest value
on every item, an event of positive
probability. To prove
(iv)$\Rightarrow$(i), we use the following bound for one-buyer menus.

\begin{lemma}[Top-value payment bound]\label{lem:top-value-payment-bound}
Suppose top selling is optimal for one buyer. Every one-buyer IC and IR
menu $(q,\pi)$ with nonnegative payments satisfies
\begin{equation}
\mathbb E_t[\pi(t)]\le
\mathbb E_t\left[
\sum_{j\in[m]}\bar v_j\one_{\{t_j=\bar v_j\}}q_j(t)
\right].
\label{eq:top-value-payment-bound}
\end{equation}
\end{lemma}

\begin{proof}
At each report $t$, increase $q_j(t)$ to one whenever $t_j=\bar v_j$,
and add
$\sum_{\substack{j\in[m]\\t_j=\bar v_j}}\bar v_j(1-q_j(t))$
to $\pi(t)$. Truthful utility is unchanged, while utility from any
deviating report weakly falls because each true value is at most
$\bar v_j$. The menu remains IC and IR with nonnegative payments.
Its revenue is at most the revenue from one-buyer top selling;
subtracting the added expected payment gives
\eqref{eq:top-value-payment-bound}.
\end{proof}

For (iv)$\Rightarrow$(i), the bound in
Lemma~\ref{lem:top-value-payment-bound} applies to each buyer's interim
menu under BIC and BIR. Buyer independence and feasibility then bound
total revenue by the expected revenue from top selling, which separate
selling attains or exceeds.
Appendix~\ref{app:reductions} gives the detailed proof.

The restriction to $n\ge2$ in the one-buyer reduction
(Theorem~\ref{thm:separate-top-single-buyer}) matters.
With one buyer, optimal separate selling need not use top prices.
See, for example, \textcite[Theorem~4.2]{MertzanidisTerzoglou2026}.

Unless stated otherwise, top-selling optimality refers to the
one-buyer problem. By Theorem~\ref{thm:separate-top-single-buyer},
top-selling optimality for one buyer implies top-selling optimality
for every $n\ge1$. We focus on the one-buyer problem
throughout the remainder of this section and
Section~\ref{sec:optimality}. In one-buyer probabilities and expectations,
$v_j$ denotes a draw from $F_j$. Dropping the buyer index, we write
\[
u(t)=\sum_{j\in[m]} t_jx_j(t)-\tau(t)
\]
for the truthful utility of type \(t\in\mathcal V\). Under top selling, type \(t\) receives exactly those items \(j\in[m]\) for which \(t_j=\bar v_j\), paying \(\bar v_j\) for each such item.

\subsection{One-buyer dual flow characterization}

We analyze the one-buyer problem through its finite-dimensional linear-programming dual.
Following \textcite[Lemma~4 and Theorem~6]{CaiDevanurWeinberg2021}, 
we interpret the dual variables as flows between types 
and use the induced virtual values to obtain an upper bound on revenue.

When the dual bound equals the revenue from top selling, complementary slackness requires every IC constraint assigned a positive dual multiplier to bind. Consider the IC constraint in which the true type is \(\hat t\) and the reported type is \(t\). Under top selling, truthful utility is zero for every type, and reporting \(t\) yields exactly the items \(j\) for which \(t_j=\bar v_j\). Hence this IC constraint binds if and only if
\[
u(\hat t)
=
u(t)+(\hat t-t)\cdot x(t) \iff
\sum_{\substack{j\in[m]\\t_j=\bar v_j}}(\bar v_j-\hat t_j)=0.
\]
Since each term in the sum is nonnegative, this is equivalent to 
requiring \(\hat t_j=\bar v_j\) whenever \(t_j=\bar v_j\). 
Thus, a dual multiplier can be positive only if 
every coordinate that is at its highest value in \(t\) 
remains at its highest value in \(\hat t\).

This observation identifies the key restriction on flows 
that can certify the optimality of top selling. 

\begin{definition}[Feasible top-preserving flow]
\label{def:top-preserving-flow}
A nonnegative flow is a function $\lambda:\mathcal V\times\mathcal V\to\mathbb R_+$, with $\lambda(t,\hat t)$ interpreted as flow from $t$ to $\hat t$.
A flow $\lambda$ is \emph{top-preserving} if, for all $t,\hat t\in\mathcal V$ and $j\in[m]$, $\lambda(t,\hat t)>0$ and $t_j=\bar v_j$ imply $\hat t_j=\bar v_j$. 
A flow $\lambda$ is \emph{feasible} if, for every $t\in\mathcal V$,
\begin{equation}
\sum_{\hat t\in\mathcal V}\lambda(\hat t,t)
-
\sum_{\hat t\in\mathcal V}\lambda(t,\hat t)
\le
\mathrm{Pr}(t),
\label{eq:flow:f1}
\end{equation}
and, for every $j\in[m]$ with $t_j<\bar v_j$,
\begin{equation}
\sum_{\hat t\in\mathcal V}
\lambda(t,\hat t)(\hat t_j-t_j)
\ge
\mathrm{Pr}(t)t_j.
\label{eq:flow:f2}
\end{equation}
\end{definition}

Condition~\eqref{eq:flow:f1} says that the total flow received at \(t\)
cannot exceed \(\mathrm{Pr}(t)\) plus the total flow sent from \(t\).
Condition~\eqref{eq:flow:f2} requires the outgoing flow from \(t\),
weighted by the change in coordinate \(j\), to be large enough to cover
the required amount \(\mathrm{Pr}(t)t_j\) whenever \(t_j<\bar v_j\). 

\begin{proposition}\label{prop:flow-characterization}
Top selling is optimal for one buyer if and only if there exists a feasible top-preserving flow $\lambda:\mathcal V\times\mathcal V\to\mathbb R_+$.
\end{proposition}
Appendix~\ref{app:flow-characterization-proof} proves the proposition.
The content beyond generic finite-dimensional duality is the
top-preserving restriction: it identifies which dual variables can be
positive at a top-selling optimum. The proposition thereby replaces
mechanism optimization with feasibility of flows satisfying
\eqref{eq:flow:f1} and \eqref{eq:flow:f2}. This test is exact for
general finite supports, although it need not yield an inequality
directly in the marginal values and probabilities.

\section{Conditions for top-selling optimality}
\label{sec:optimality}

The previous section provides an exact characterization of top-selling
optimality in terms of the existence of a feasible top-preserving flow.
However, checking flow feasibility requires solving a linear program whose
size grows rapidly with the number of items and support values. We therefore
seek conditions stated directly in terms of the marginal distributions that
are easier to verify.

We first derive a necessary condition, called the \emph{top--tail revenue
bound} (TTRB), which compares the potential revenue gains from selling
at non-top valuations with the revenue loss at the highest type. We
establish this bound using deviations that add a discounted grand bundle
and item lotteries to top selling. We then show that TTRB exactly
characterizes flow feasibility on types with at most one item below its
highest value. An example shows why TTRB alone does not establish optimality:
a bundle can earn more than top selling even when the simple
deviations captured by TTRB do not. We then characterize when an upward
flow certifies optimality. Myerson regularity and boundary dominance
each identify settings in which TTRB is also sufficient.
Proofs omitted from this section are in Appendix~\ref{app:flow-proofs}.

\subsection{The Top--Tail Revenue Bound}
\label{sec:necessary-condition}

For each item \(j\in[m]\) and \(1\le k<K_j\), define
\[
\eta_{j,k}:=
\frac{\sum_{\ell=k}^{K_j-1}p_j(v_{j,\ell})v_{j,k}}
{p_j(\bar v_j)(\bar v_j-v_{j,k})}.
\]
Equivalently, the ratio can be written as
\[
\eta_{j,k}
=
\frac{
\mathrm{Pr}(v_{j,k}\le v_j<\bar v_j)\,v_{j,k}
}{
\mathrm{Pr}(v_j=\bar v_j)\,(\bar v_j-v_{j,k})
}.
\]
Thus, the ratio is large when there is substantial probability
mass on non-top values at or above \(v_{j,k}\), while the highest
value is unlikely or not much higher than \(v_{j,k}\).
\begin{definition}[Top--Tail Revenue Bound]
\label{def:ttrb}
We say that $\mathbf F$ satisfies TTRB if
\begin{equation}
\sum_{j\in[m]}\max_{1\le k<K_j}\eta_{j,k}\le1.
\label{eq:ttrb}
\end{equation}
\end{definition}

The following menu deviation explains the economic content of TTRB.
For each item $j$,
choose $k_j$ attaining $\max_{1\le k<K_j}\eta_{j,k}$, and take a positive
$\varepsilon$ smaller than every gap
$\bar v_j-v_{j,K_j-1}$. Add to the one-buyer top-selling menu a grand
bundle at price $\sum_{j\in[m]}\bar v_j-\varepsilon$. For each item $j$,
also offer every other item with certainty and item $j$ with probability
$\varepsilon/(\bar v_j-v_{j,k_j})$, at price
\[
\sum_{h\in[m]\setminus\{j\}}\bar v_h
+\frac{\varepsilon v_{j,k_j}}{\bar v_j-v_{j,k_j}}.
\]
Resolve indifference in favor of the option with the highest payment and,
if payments also tie, in favor of an added option.
The type $\bar v$ buys the grand bundle, reducing revenue there by
$\varepsilon$. A type with exactly item $j$ below its highest value
chooses the added lottery precisely when its value for that item is at
least $v_{j,k_j}$; the lottery raises payment by
$\varepsilon v_{j,k_j}/(\bar v_j-v_{j,k_j})$.
All other types retain their top-selling choice. Hence the exact change
in expected revenue is
\[
\varepsilon\mathrm{Pr}(\bar v)
\left(\sum_{j\in[m]}\eta_{j,k_j}-1\right).
\]
Thus, TTRB compares the revenue gained from selling below the highest
value at these types with the price reduction at $\bar v$.

\begin{proposition}\label{prop:ttrb-necessary}
If $M^{\mathrm{top}}$ is optimal for one buyer, then TTRB, \eqref{eq:ttrb}, holds.
\end{proposition}
The preceding menu deviation proves the proposition.

TTRB imposes a quantitative restriction on the probability
of high non-top valuations. For every $\alpha\in(0,1)$,
\[
\sum_{j\in[m]}
\frac{
\mathrm{Pr}(\alpha\bar v_j\le v_j<\bar v_j)
}{
p_j(\bar v_j)
}
\le \frac{1-\alpha}{\alpha}.
\]
To see this, for each item choose the smallest non-top
support value at least $\alpha\bar v_j$, if one exists.
The corresponding TTRB ratio is at least
$\alpha/(1-\alpha)$ times the probability ratio appearing
above. Summing across items gives the inequality.
Thus, valuations close to the highest value cannot carry
substantial probability mass relative to the top, and
these restrictions accumulate across items. This is a
necessary restriction, not a sufficient condition for
top-selling optimality.

TTRB exactly characterizes the feasibility of a top-preserving flow
on the restricted type space consisting of types with at most one item
below its highest value.
Proposition~\ref{prop:boundary-flow} in Appendix~\ref{app:flow-proofs}
gives the formal statement and proof.

TTRB is not sufficient for top-selling optimality, as the following
example shows.

\begin{example}[TTRB is not sufficient]
\label{ex-ttrb-not-sufficient}
Let $m=2$ and suppose the two items are i.i.d.\ with common support
$V=\left\{\frac15,\frac35,1\right\}$. For each $j\in[2]$, let
$p_j(v_{j,1})=\frac{11}{20}$, $p_j(v_{j,2})=\frac{1}{10}$, and
$p_j(\bar v_j)=\frac{7}{20}$.
For either item, direct substitution gives
\[
\eta_{j,1}=\frac{13}{28}>\eta_{j,2}=\frac37.
\]
Hence \(\sum_{j\in[2]}\max_{1\le k<K_j}\eta_{j,k}=13/14<1\),
so TTRB holds.

For either item, posting prices $1$, $3/5$, and $1/5$ yields expected
revenue $7/20$, $27/100$, and $1/5$, respectively. Thus separate selling
coincides with top selling and earns $7/10$. If the buyer purchases when
indifferent, selling the grand bundle at price \(6/5\) makes a sale
whenever at least one item has value \(1\), and also at
\((3/5,3/5)\). Its expected revenue is
\[
\frac65\left[1-\left(\frac{13}{20}\right)^2
+\left(\frac1{10}\right)^2\right]
=\frac{141}{200}>\frac7{10}.
\]
Thus TTRB holds although top selling is not optimal. \hfill$\triangle$
\end{example}

A natural competitor to top selling is a \emph{discounted grand
bundle}. In this example, it extracts $6/5$ from $(3/5,3/5)$, which
buys neither item under top selling. It also raises the payment from
$1$ to $6/5$ when exactly one item has value $1$, while reducing the
payment from $2$ to $6/5$ at $(1,1)$. The additional revenue from
types such as $(3/5,3/5)$, together with the higher payments when exactly
one item has value $1$, can outweigh that loss even when TTRB holds.
Reducing the probability of such types can make discounted bundling less
profitable and may restore top-selling optimality. However, TTRB alone does
not rule out profitable deviations involving types with multiple non-top
values. We therefore seek additional conditions on the value distributions
that can guarantee top-selling optimality.

\subsection{The slice-tail condition}
\label{sec:upward}

We next ask when the flow characterization can certify optimality
using only movements toward higher types. For each item \(j\in[m]\) and \(1\le k<K_j\), define
\[
\rho_{j,k}:=
\frac{\sum_{\ell=k}^{K_j-1}p_j(v_{j,\ell})v_{j,\ell}}
{\sum_{\ell=k+1}^{K_j}p_j(v_{j,\ell})(v_{j,\ell}-v_{j,k})}.
\]
The ratio is large when there is substantial probability
mass on non-top values at or above \(v_{j,k}\),
while values above \(v_{j,k}\) are not much higher than \(v_{j,k}\).
Every denominator in $\rho_{j,k}$ is at least
$p_j(\bar v_j)(\bar v_j-v_{j,k})>0$, so the ratios are well-defined.
The ratios are nonnegative. For item $j$, all of them are zero precisely
when $V_j=\{0,\bar v_j\}$.
\begin{definition}[Slice-tail condition]
\label{def:slice-tail}
We say that $\mathbf F$ satisfies the \emph{slice-tail condition} if
\begin{equation}
\sum_{j\in[m]}\max_{1\le k<K_j}\rho_{j,k}\le1.
\label{eq:slice-tail}
\end{equation}
\end{definition}

To connect the slice-tail condition to the flow problem, we consider flows that move only toward higher types.
\begin{definition}[Upward flow]
\label{def:upward-flow}
A flow $\lambda$ is \emph{upward} if $\lambda(t,\hat t)>0$ implies $\hat t_j\ge t_j$
for all $t,\hat t\in\mathcal V$ and $j\in[m]$.
\end{definition}

Every upward flow is top-preserving: \(t_j=\bar v_j\) and
\(\lambda(t,\hat t)>0\) imply \(\hat t_j=\bar v_j\).

\begin{proposition}\label{prop:upward}
A feasible upward flow $\lambda:\mathcal V\times\mathcal V\to\mathbb R_+$
exists if and only if $\mathbf F$ satisfies the slice-tail condition
\eqref{eq:slice-tail}.
\end{proposition}

Thus, \eqref{eq:slice-tail} characterizes exactly when an upward
flow can certify top-selling optimality.

\begin{corollary}\label{cor:slice-tail-sufficient}
If $\mathbf F$ satisfies the slice-tail condition, then top selling is optimal.
\end{corollary}

\begin{proof}
Proposition~\ref{prop:upward} gives a feasible upward flow.
Every upward flow is top-preserving, so
Proposition~\ref{prop:flow-characterization} implies top-selling optimality.
\end{proof}

If top selling is optimal while \eqref{eq:slice-tail} fails,
any feasible top-preserving flow must have
\(\lambda(t,\hat t)>0\) with \(\hat t_j<t_j\) for some types
\(t,\hat t\in\mathcal V\) and item \(j\in[m]\).
As shown below, top selling can indeed be optimal even when no feasible
upward flow exists.

We first compare the slice-tail condition with TTRB. The slice-tail
condition implies TTRB.
To see why, fix \(j\in[m]\)
and \(1\le k<K_j\).
The slice-tail ratio $\rho_{j,k}$ can be written as
\[
\frac{
v_{j,k}\mathrm{Pr}(v_{j,k}\le v_j<\bar v_j)
+\mathbb E\left[(v_j-v_{j,k})
\one_{\{v_{j,k}\le v_j<\bar v_j\}}\right]
}{
p_j(\bar v_j)(\bar v_j-v_{j,k})
+\mathbb E\left[(v_j-v_{j,k})
\one_{\{v_{j,k}\le v_j<\bar v_j\}}\right]
}.
\]
The expectation collects the amounts by which values below $\bar v_j$
exceed $v_{j,k}$.
By comparison, the ratio $\eta_{j,k}$ in the definition of TTRB is
\[
\frac{
v_{j,k}\mathrm{Pr}(v_{j,k}\le v_j<\bar v_j)
}{
p_j(\bar v_j)(\bar v_j-v_{j,k})
}.
\]
The slice-tail ratio therefore adds the same nonnegative expectation to
the numerator and denominator of the TTRB ratio.
Under \eqref{eq:slice-tail}, every $\rho_{j,k}\le1$. Removing the common
nonnegative expectation therefore cannot increase the ratio, so
$\eta_{j,k}\le\rho_{j,k}$ for every $j\in[m]$ and $1\le k<K_j$.
Consequently,
\[
\sum_{j\in[m]}\max_{1\le k<K_j}\eta_{j,k}
\le\sum_{j\in[m]}\max_{1\le k<K_j}\rho_{j,k}
\le1,
\]
proving that the slice-tail condition implies TTRB.
Example~\ref{ex-ttrb-not-sufficient} shows that the converse fails.

The extra expectation in \(\rho_{j,k}\) accounts for values strictly
between \(v_{j,k}\) and \(\bar v_j\). It vanishes when no such
value has positive probability. This explains why TTRB and the
slice-tail condition coincide for binary supports but may differ
when an item has intermediate values.

\subsection{When is TTRB sufficient?}
\label{sec:ttrb-sufficient}

In this subsection, we give two conditions under which TTRB is also sufficient
for top-selling optimality.
The first is standard \emph{Myerson regularity}.
The second is the \emph{boundary-dominance condition}, which requires each
item's second-highest value to be sufficiently small relative to its highest
value, with a bound determined by the probabilities of top values across items.

For $j\in[m]$ and $k\in[K_j]$, write
\[
T_{j,k}:=\sum_{\ell=k}^{K_j}p_j(v_{j,\ell}).
\]
For each item $j\in[m]$ and $1\le k\le K_j-1$, the \emph{virtual value}
at $v_{j,k}$ is
\[
\phi_{j,k}:=v_{j,k}-\frac{T_{j,k+1}}{p_j(v_{j,k})}
(v_{j,k+1}-v_{j,k}).
\]
At $k=K_j$, it is $\phi_{j,K_j}:=\bar v_j$.
The marginal distribution $F_j$ is
\emph{Myerson regular} if
\[
\phi_{j,1}\le\phi_{j,2}\le\cdots\le\phi_{j,K_j}.
\]
\begin{proposition}[Myerson regularity]\label{prop:regularity}
Suppose $F_j$ is Myerson regular for every $j\in[m]$. Then top selling is optimal if and only if
\begin{equation}
\sum_{j\in[m]}
\frac{p_j(v_{j,K_j-1})v_{j,K_j-1}}
{p_j(\bar v_j)(\bar v_j-v_{j,K_j-1})}\le1.
\label{eq:regular-terminal}
\end{equation}
When \eqref{eq:regular-terminal} holds, a feasible upward flow exists.
\end{proposition}

Condition~\eqref{eq:regular-terminal} is precisely
the part of TTRB obtained by evaluating \(\eta_{j,k}\) at \(k=K_j-1\)
for each item \(j\). Thus, under Myerson regularity, TTRB is sufficient
as well as necessary for top-selling optimality.
Necessity follows from TTRB. For sufficiency, Myerson regularity makes
\eqref{eq:regular-terminal} imply the slice-tail condition, so
Corollary~\ref{cor:slice-tail-sufficient} applies.
For identical binary items with support $\{a,b\}$ and probability
$q$ at $b$, condition~\eqref{eq:regular-terminal} becomes
$m(1-q)a\le q(b-a)$. For two items, this recovers the optimality
region in \textcite{Yao2017}. For regular distributions with more than
two support values, the
analogous condition depends only on the two highest values of each
item and their respective probabilities.

We next consider boundary dominance, which makes TTRB sufficient without
Myerson regularity.
\begin{definition}[Boundary dominance]
\label{def:boundary-dominance}
We say that $\mathbf F$ satisfies \emph{boundary dominance} if
\begin{equation}
\frac{v_{j,K_j-1}}{\bar v_j}
\le\frac{\mathrm{Pr}(\bar v)}{\max_{k\in[m]\setminus\{j\}}p_k(\bar v_k)}
\label{eq:primitive-bd:pbd}
\end{equation}
for every item $j\in[m]$.
\end{definition}

The right-hand side of \eqref{eq:primitive-bd:pbd} equals
\[
\frac{\mathrm{Pr}(\bar v)}
{\max_{k\in[m]\setminus\{j\}}p_k(\bar v_k)}
=
\min_{k\in[m]\setminus\{j\}}
\mathrm{Pr}\!\left(v_\ell=\bar v_\ell\ \text{for all }\ell\in[m]\setminus\{k\}\right).
\]
Boundary dominance is therefore easier to satisfy when the
second-highest value of each item is small relative to its highest
value and the highest values occur with high probability. When the
item values are i.i.d.,
the condition reduces to
\[
\frac{v_{K-1}}{\bar v}\le p(\bar v)^{m-1}.
\]
This expression also shows that the condition becomes
more restrictive as the number of items increases.

\begin{proposition}[Boundary dominance]
\label{prop:boundary-dominance}
Suppose TTRB, \eqref{eq:ttrb}, and boundary dominance,
\eqref{eq:primitive-bd:pbd}, hold. Then top selling is optimal for one buyer.
\end{proposition}

The proof first uses TTRB to construct a flow on types with at most one item
below its highest value. It then uses the residual capacity at these types
under \eqref{eq:flow:f1} to receive flow from types with at least two items
below their highest values. The added flow preserves every coordinate of its
source type already at its highest value, while boundary dominance ensures
that \eqref{eq:flow:f2} holds for these source types.

As under Myerson regularity, TTRB exactly characterizes top-selling
optimality when boundary dominance holds. However, the slice-tail condition
and the conjunction of TTRB with boundary dominance are logically
independent. In particular, top selling can be optimal even when no feasible
upward flow exists, as illustrated by
Example~\ref{ex:sufficient-conditions-independent} in
Appendix~\ref{app:flow-proofs}.

\section{Conclusion}
\label{sec:conclusion}

Our results show that the exact optimality of separate selling is highly
restrictive. With at least two buyers, and under our standing support
assumption, separate selling is optimal if and only if top selling is optimal
for a single buyer, under both DSIC and BIC. Our distributional results reveal
a common economic feature behind this restriction: \emph{top-value isolation}.
High non-top valuations must be sufficiently far below the highest value or
sufficiently rare relative to it. This does not require binary supports.
Rather, it requires valuation gaps and probability masses near the top to
jointly limit the profitability of selling below the highest value.

TTRB provides a necessary quantitative expression of top-value isolation, and
under Myerson regularity, a simple inequality involving only the two highest
values becomes necessary and sufficient. Beyond regularity, top-preserving
flows characterize the full optimality region, while the slice-tail condition
identifies precisely when upward flows suffice. Although more general flows
may be needed, the same necessary restrictions on high non-top valuations
continue to apply. Thus, increasing competition may make separate selling
approximately optimal, but it does not enlarge the class of distributions for
which separate selling is exactly optimal once there are at least two buyers.

\appendix
\section{Benchmarks and proofs}

\subsection{Single-item benchmarks}
\label{sec:single-item}

Throughout this subsection, fix one item and suppress its item index. 
Write $V,F,K,\bar v,v_k,p,T_k$ for
$V_j,F_j,K_j,\bar v_j,v_{j,k},p_j,T_{j,k}$.
Set $T_{K+1}:=0$.
We use the finite-support version of the optimal-auction characterization
of \textcite{Myerson1981}.\footnote{See \textcite{Elkind2007} and
\textcite{GiannakopoulosHahn2026} for discrete treatments of the payment
formula and ironing.} We record the notation and standard conclusions
needed below.

In this notation, the virtual values defined in
Section~\ref{sec:ttrb-sufficient} are
\[
\phi_k=v_k-\frac{T_{k+1}}{p(v_k)}(v_{k+1}-v_k)
\]
for $k\in[K-1]$, with $\phi_K=\bar v$. The revenue curve $R:[0,1]\to\mathbb R$
is the piecewise-linear function through $R(0)=0$ and $R(T_k)=v_kT_k$
for $k\in[K]$. The slope of $R$ on $[T_{k+1},T_k]$ is $\phi_k$.
Let $\bar R:[0,1] \to \mathbb R$ be the least concave majorant of $R$. The ironed virtual
values are
\[
\bar\phi_k:=\frac{\bar R(T_k)-\bar R(T_{k+1})}{p(v_k)}
\]
for $k\in[K]$, and satisfy
$\bar\phi_1\le\cdots\le\bar\phi_K$.
Write $\bar\phi(v_k)=\bar\phi_k$.
Restoring the item index gives $\bar\phi_j(v_{j,k})=\bar\phi_{j,k}$
and the revenue curves $R_j,\bar R_j$.
An \emph{ironing block} is a set $\{v_a,\ldots,v_b\}$, where
$1\le a\le b\le K$, such that $\bar\phi_a=\cdots=\bar\phi_b$,
$\bar\phi_{a-1}<\bar\phi_a$ if $a>1$, and
$\bar\phi_b<\bar\phi_{b+1}$ if $b<K$. An ironing block is a 
maximal set of consecutive support values that share the same ironed virtual value.

For a monotone allocation $0\le x_1\le\cdots\le x_K\le1$, the
largest payments consistent with IC and IR are
\begin{equation}
\tau_k=v_kx_k-\sum_{r\in[k-1]}(v_{r+1}-v_r)x_r
\label{eq:single-item-payments}
\end{equation}
for $k\in[K]$, where the sum is zero when $k=1$. These payments are
nonnegative and give the lowest type $v_1$ zero utility. For any one-buyer IC
and IR menu $(x_k,\tau_k)_{k\in[K]}$,
even with \emph{signed payments},\footnote{Signed payments may take values
in $\mathbb R$. Here $\tau_k>0$ is paid by the buyer to the seller, whereas
$\tau_k<0$ is a transfer of $-\tau_k$ from the seller to the buyer.}
the payment formula and ironing give
\begin{equation}
\sum_{k\in[K]}p(v_k)\tau_k
\le\sum_{k\in[K]}p(v_k)\phi_kx_k
\le\sum_{k\in[K]}p(v_k)\bar\phi_kx_k.
\label{eq:single-item-revenue-bound}
\end{equation}
Payments given by \eqref{eq:single-item-payments} attain the first
inequality and the second is an equality 
whenever the allocation is constant within each ironing block.

In the Myerson auction described below, each buyer's payment is determined
by \eqref{eq:single-item-payments}, holding the other buyers' reports fixed.

\begin{lemma}[Myerson]
\label{thm:single-item-myerson}
For $m=1$, an optimal DSIC and IR mechanism allocates the item uniformly
among buyers with the highest strictly positive ironed virtual value,
leaving it unallocated if no such value exists.
The optimal revenue is
\[
\Rev_n(F)
=\mathbb E\left[\max\{0,\bar\phi(v_{11}),\ldots,\bar\phi(v_{n1})\}\right].
\]
\end{lemma}

We next use this characterization to determine exactly 
when top selling is optimal in the single-item problem. 

\begin{proposition}
\label{prop:single-item-top}
For $m=1$ and any $n\ge1$, top selling is optimal if and only if
$\bar\phi_k\le0$ for every $k\in[K-1]$.
\end{proposition}

\begin{proof}[Proof of Proposition~\ref{prop:single-item-top}]
For every $k\in[K-1]$, $R(T_k)=v_kT_k<\bar v T_k$. 
Then the revenue curve satisfies $R(q)\le\bar v q$ for every $q\in[0,1]$,
with equality at $q=0$ and $q=p(\bar v)$. 
The function $\bar v q$ is itself a concave majorant of $R$, so
$R(q)\le\bar R(q)\le\bar v q$, while concavity of $\bar R$ and the
resulting equality at $0$ and $p(\bar v)$ force
$\bar R(q)\ge\bar v q$ between these points.
Hence $\bar R(q)=\bar v q$ on $[0,p(\bar v)]$, and every linear segment
of $\bar R$ on $[p(\bar v),1]$ has slope strictly less than $\bar v$.
Thus, the highest support value $\bar v$ forms a singleton ironing block.

If $\bar\phi_k\le0$ for every $k\in[K-1]$, the only types with positive 
ironed virtual value are those reporting $\bar v$, whose ironed virtual 
value equals $\bar v$. Hence, at every profile, the allocation rule of 
top selling maximizes ironed virtual surplus. 
Lemma~\ref{thm:single-item-myerson} therefore implies that top selling is optimal.

Conversely, suppose $\bar\phi_k>0$ for some $k\in[K-1]$. When all buyers
report $v_k$, an event with positive probability, top selling allocates
nothing, whereas the maximum ironed virtual surplus is strictly positive.
At every profile, the maximum ironed virtual surplus is at least that
generated by top selling, which equals its revenue. Hence, by
Lemma~\ref{thm:single-item-myerson}, the optimal expected revenue is strictly
greater than that of top selling.
\end{proof}

The standing support assumption rules out a class of environments in which
separate selling is optimal under DSIC and IR for every $n\ge1$.

\begin{lemma}\label{lem:degenerate-items-separate}
If all but at most one item have support $\{0,\bar v_j\}$, separate
selling is optimal under DSIC and IR for every $n\ge1$.
\end{lemma}

\begin{proof}
Choose item $h$ so that every item $j\in[m]\setminus\{h\}$ has
support $\{0,\bar v_j\}$; if every item has this support, choose
any $h\in[m]$. Fix the reported values of all items other than $h$.
For any feasible DSIC and IR mechanism $(x,\tau)$, give buyer $i$ the
same allocation probability $x_{ih}$ for item $h$ and the adjusted
payment
\[
\tau_i-\sum_{j\in[m]\setminus\{h\}}v_{ij}x_{ij}.
\]
As functions of the item-$h$ reports, these rules form a feasible
single-item DSIC and IR mechanism. The adjusted payments may be
negative, but the single-item revenue bound
\eqref{eq:single-item-revenue-bound} also applies to signed payments.
Apply that bound to each buyer's conditional item-$h$ menu and use
feasibility, as in Lemma~\ref{thm:single-item-myerson}.
Independence of values across items then bounds the adjusted
payments' conditional expected sum by $\Rev_n(F_h)$.

For every item $j\ne h$, feasibility and
$v_{ij}\in\{0,\bar v_j\}$ imply
\[
\sum_{i\in[n]}v_{ij}x_{ij}(v)
\le
\bar v_j\one_{\{v_{ij}=\bar v_j\text{ for some }i\in[n]\}}.
\]
Taking expectations and adding the adjusted-payment bound gives
\[
\Rev_n(\mathbf F)
\le\Rev_n(F_h)
+\sum_{j\in[m]\setminus\{h\}}
\bar v_j\mathrm{Pr}(v_{ij}=\bar v_j\text{ for some }i\in[n]).
\]
For each $j\ne h$, selling item $j$ at $\bar v_j$ earns its full
available surplus, so the right-hand side equals
$\SREV_n(\mathbf F)$. Separate selling attains this bound.
\end{proof}

\subsection{Proof of the one-buyer reduction}
\label{app:reductions}

\begin{proof}[Proof of Theorem~\ref{thm:separate-top-single-buyer}]
We prove (i)$\Rightarrow$(ii)$\Rightarrow$(iii)$\Rightarrow$(iv)$\Rightarrow$(i).
The implication (ii)$\Rightarrow$(iii) uses perturbations of the product
of optimal single-item auctions.

\emph{(i)$\Rightarrow$(ii).}
Since $\SREV_n(\mathbf F)\le\Rev_n(\mathbf F)\le\Rev^B_n(\mathbf F)$,
(i) implies (ii).

\emph{(ii)$\Rightarrow$(iii).}
Suppose separate selling is optimal under DSIC and IR. By the standing
support assumption,
at least two items have a strictly positive non-top support value, i.e., 
\(v_{j,K_j-1}>0\) for at least two \(j\in[m]\).

For each item, run the Myerson auction described in 
Lemma~\ref{thm:single-item-myerson}, with uniform 
tie-breaking and canonical payments.
The product of these single-item auctions attains
\(\SREV_n(\mathbf F)\). Since separate selling is optimal by hypothesis,
this product mechanism is optimal.

We show that top selling must be optimal for every item. Suppose, toward a
contradiction, that top selling is suboptimal for at least one item. For each
such item $j\in[m]$, let $B_j$ be the highest ironing block strictly below
$\bar v_j$ with strictly positive ironed virtual value, and put
$\ell_j:=\min B_j$.
Such a block exists by Proposition~\ref{prop:single-item-top}.
Since the ironed virtual values are nondecreasing, \(B_j\) contains
\(v_{j,K_j-1}\). 

Moreover, \(\ell_j>0\). Indeed, if \(v_{j,1}=0\), then 
$\bar R_j(T_{j,1})=R_j(T_{j,1})=v_{j,1}T_{j,1}=0$ and
$\bar R_j(T_{j,2})\ge R_j(T_{j,2})=v_{j,2}T_{j,2}>0$.
Hence the ironed virtual value at \(v_{j,1}=0\) is 
\[\bar\phi_{j,1}=\frac{\bar R_j(T_{j,1})-\bar R_j(T_{j,2})}{p_j(v_{j,1})}<0.\]
Therefore, zero cannot belong to $B_j$, so $\ell_j>0$.

The highest support value $\bar v_j$ forms a singleton ironing block.
Fix the reports of buyers $3,\ldots,n$ at their lowest support values for
item $j$. Let $q_j(v,w)$ denote buyer $1$'s allocation probability in the
Myerson auction for item $j$ when buyers $1$ and $2$ report $v,w\in V_j$,
respectively. By symmetry, buyer $2$'s allocation probability is $q_j(w,v)$.
Define
\[
\begin{aligned}
\sigma_j(v,w)&:=q_j(v,w)-q_j(w,v),\\
f_j(v)&:=\frac{(v-\ell_j)_+}{\bar v_j-\ell_j}.
\end{aligned}
\]

\begin{claim}\label{clm:reduction-scalar-comparisons}
Fix an item $j$ for which top selling is suboptimal, and let
$v,\hat v,w\in V_j$. If the IC constraint with true value $v$ and report
$\hat v$ binds in the Myerson auction for item $j$, then
\begin{equation}
\sigma_j(v,w)\ge\sigma_j(\hat v,w).
\label{eq:reduction-scalar-sigma}
\end{equation}
If \(w\in B_j\), every such binding constraint also satisfies
\begin{equation}
f_j(v)-f_j(\hat v)
=
\frac{v-\hat v}{\bar v_j-\ell_j}
\one_{\{\hat v\in B_j\}},
\label{eq:reduction-scalar-f}
\end{equation}
and
\begin{equation}
\begin{aligned}
\hat v=\bar v_j&\Longrightarrow v=\bar v_j,\\
v=\bar v_j&\Longrightarrow\hat v\in B_j\cup\{\bar v_j\}.
\end{aligned}
\label{eq:reduction-scalar-top-reports}
\end{equation}
If \(w\in B_j\), the allocation rule also gives
\begin{equation}
\begin{aligned}
\sigma_j(\hat v,w)&=0 &&\text{for every }\hat v\in B_j,\\
\sigma_j(\bar v_j,w)&=1.
\end{aligned}
\label{eq:reduction-scalar-block-sigma}
\end{equation}
\end{claim}

\begin{proof}[Proof of Claim~\ref{clm:reduction-scalar-comparisons}]
The function \(\sigma_j\) is antisymmetric in the sense that
\[\sigma_j(w,v)=q_j(w,v)-q_j(v,w)=-\sigma_j(v,w)\]
and nondecreasing in its first argument. 
This gives \eqref{eq:reduction-scalar-sigma} 
when \(\hat v\le v\). 
If instead \(\hat v>v\), write \(v=v_{j,k}<\cdots<v_{j,\ell}=\hat v\), and
let \(x_r\) denote buyer \(1\)'s allocation probability at \(v_{j,r}\).
By IC, \(x_k\le\cdots\le x_\ell\), while the canonical payment 
formula gives \(u(\hat v)-u(v)=\sum_{r=k}^{\ell-1}(v_{j,r+1}-v_{j,r})x_r\).
If the IC constraint with true value \(v\) and report \(\hat v\) binds,
then \(u(\hat v)-u(v)=(\hat v-v)x_\ell\). 
Hence \(\sum_{r=k}^{\ell-1}(v_{j,r+1}-v_{j,r})(x_\ell-x_r)=0\).
Since every term is nonnegative, \(x_r=x_\ell\) for all \(r=k,\ldots,\ell\), 
so buyer \(1\)'s allocation probability is constant over the support 
values between \(v\) and \(\hat v\).
Since the other buyers' reports are fixed and the auction ranks buyers by
ironed virtual values with uniform tie-breaking, buyer $1$'s constant
allocation probability implies that buyer $2$'s allocation probability is
also constant over this range. Hence
$\sigma_j(v,w)=\sigma_j(\hat v,w)$, proving \eqref{eq:reduction-scalar-sigma}.

When $w\in B_j$, buyer $1$'s allocation probability is zero below $\ell_j$,
equals $q_j(\ell_j,w)\in(0,1)$ throughout $B_j$, and equals one at $\bar v_j$.
By the canonical payment formula, her utility is zero below $\ell_j$ and
equals $q_j(\ell_j,w)(v-\ell_j)$ for $v\in B_j$. At the top value, her utility
is $q_j(\ell_j,w)(\bar v_j-\ell_j)$, since the increase in allocation
probability at $\bar v_j$ creates no additional information rent.

We now characterize the binding IC constraints. Suppose $\hat v\in B_j$.
Since $q_j(\hat v,w)=q_j(\ell_j,w)$ and
$u(\hat v)=q_j(\ell_j,w)(\hat v-\ell_j)$, the IC constraint with true value
$v$ and report $\hat v$ binds if and only if
\[
u(v)=q_j(\ell_j,w)(v-\ell_j).
\]
By the utility formula above, this equality holds precisely when
$v\in B_j\cup\{\bar v_j\}$. In either case,
\[
f_j(v)-f_j(\hat v)
=
\frac{v-\hat v}{\bar v_j-\ell_j}.
\]

If instead \(\hat v<\ell_j\), then \(q_j(\hat v,w)=u(\hat v)=0\). 
Binding therefore requires \(u(v)=0\). 
Since \(u(v)>0\) whenever \(v>\ell_j\), this implies 
\(v\le\ell_j\), and hence \(f_j(v)=f_j(\hat v)=0\). 
Combining the two cases gives
\[
f_j(v)-f_j(\hat v)
=
\frac{v-\hat v}{\bar v_j-\ell_j}
\one_{\{\hat v\in B_j\}},
\]
provided \(\hat v\neq\bar v_j\).

Finally, suppose \(\hat v=\bar v_j\). Since \(q_j(\bar v_j,w)=1\), binding requires
\[
u(v)-u(\bar v_j)=v-\bar v_j.
\]
If \(v\in B_j\), the left-hand side equals \(q_j(\ell_j,w)(v-\bar v_j)\), 
which is strictly greater than \(v-\bar v_j\) because \(q_j(\ell_j,w)<1\) and \(v<\bar v_j\). 
If $v<\ell_j$, the left-hand side equals $-q_j(\ell_j,w)(\bar v_j-\ell_j)$,
which is strictly greater than $v-\bar v_j$ since $q_j(\ell_j,w)<1$ and
$v<\ell_j$.
Thus binding is possible only when \(v=\bar v_j\).
Conversely, suppose $v=\bar v_j$. A report $\hat v<\ell_j$ yields zero utility
and cannot bind because $u(\bar v_j)>0$. For any $\hat v\in B_j$,
\[
u(\hat v)+(\bar v_j-\hat v)q_j(\hat v,w)
=
q_j(\ell_j,w)(\bar v_j-\ell_j)
=
u(\bar v_j).
\]
The truthful report also binds. This proves
\eqref{eq:reduction-scalar-top-reports}.

Equal treatment within
\(B_j\) and the separate ironing block at \(\bar v_j\) give
\eqref{eq:reduction-scalar-block-sigma}.
\end{proof}

We now consider two cases.

\medskip
\noindent
\emph{Case 1: top selling is suboptimal for at least two items.}

Choose two such items and relabel them \(1\) and \(2\). Let \(E\) be the
event that buyers \(3,\ldots,n\) report their lowest support values on
these two items. If \(n=2\), this event always occurs.
Since all support probabilities are strictly positive, $\mathrm{Pr}(E)>0$. 
The event \(E\) is independent of the reports of buyers \(1\) and \(2\).

Write \(t \in \mathcal V\) and \(s \in \mathcal V\) for the types of buyers \(1\) and \(2\).
For \(\varepsilon>0\), to be chosen sufficiently small, modify 
only these two buyers and only on \(E\). For buyer \(1\) and
\(j,j'\in\{1,2\}\) with \(j\ne j'\), set
\begin{equation}
\Delta x_{1j}(t,s)
=
\frac{\varepsilon}{\bar v_j-\ell_j}
\one_{\{t_j,s_j\in B_j\}}
\sigma_{j'}(t_{j'},s_{j'}),
\label{eq:reduction-two-items-allocation}
\end{equation}
and
\begin{equation}
\begin{aligned}
\frac{\Delta u_1(t,s)}{\varepsilon}
&=
\sum_{\substack{j,j'\in\{1,2\}\\j\ne j'}}
\one_{\{s_j\in B_j\}}
f_j(t_j)\sigma_{j'}(t_{j'},s_{j'})
\\
&-\one_{\{s_1\in B_1,s_2\in B_2,t_1=\bar v_1,t_2=\bar v_2\}}.
\end{aligned}
\label{eq:reduction-two-items-utility}
\end{equation}
Define buyer \(2\)'s changes symmetrically by interchanging \(t\) and \(s\) in the formulas above.
For example, $\Delta x_{21}(t,s)=\frac{\varepsilon}{\bar v_1-\ell_1}\one_{\{t_1,s_1\in B_1\}}
\sigma_2(s_2,t_2)$.
Set \(\Delta x_{ij}(v)=0\) for all other buyer-item pairs and on \(E^c\),
and set \(\Delta u_i(v)=0\) for buyers \(i\ge3\) and on \(E^c\).
For every profile $v$, define the perturbed allocations and utilities by
$x'_{ij}(v):=x_{ij}(v)+\Delta x_{ij}(v)$ for every $i\in[n]$ and $j\in[m]$,
and $u_i'(v):=u_i(v)+\Delta u_i(v)$ for every $i\in[n]$.
Given \(x'\) and \(u'\), define the corresponding payments by
\[
\tau_i'(v)
:=
\sum_{j\in[m]}v_{ij}x_{ij}'(v)-u_i'(v).
\]

By \eqref{eq:reduction-two-items-allocation} and antisymmetry of \(\sigma_j\),
the changes in allocation probabilities 
sum to zero across buyers \(1\) and \(2\) 
for each affected item, 
so the total allocation of that item is unchanged. 
Moreover, allocations are modified only when both buyers have strictly
positive probabilities of receiving the item. Thus, for sufficiently small
$\varepsilon>0$, the modified probabilities remain nonnegative. Since their
sum is unchanged, feasibility is preserved.

Fix the other buyers' reports so that $E$ holds. Consider a binding DSIC
constraint for buyer $1$, with true type $t\in\mathcal V$ and report
$\hat t\in\mathcal V$. Since the original mechanism sells items separately,
the DSIC slack is the sum of the corresponding single-item IC slacks. Each is
nonnegative, so every corresponding single-item IC constraint must bind.

Since the original DSIC slack is zero, its slack after the perturbation is
\[
u_1'(t,s)-u_1'(\hat t,s)
-(t-\hat t)\cdot x_1'(\hat t,s)
=\Delta u_1(t,s)-\Delta u_1(\hat t,s)
-(t-\hat t)\cdot\Delta x_1(\hat t,s).
\]
Only items \(1\) and \(2\) are affected.

Substituting the definitions of \(\Delta u_1\) and \(\Delta x_1\), 
and dividing by \(\varepsilon\), gives
\[
\begin{aligned}
&\frac{1}{\varepsilon}
\left[
\Delta u_1(t,s)-\Delta u_1(\hat t,s)
-(t-\hat t)\cdot\Delta x_1(\hat t,s)
\right]
\\
={}&
\one_{\{s_1\in B_1\}}
\left[
f_1(t_1)\sigma_2(t_2,s_2)
-
f_1(\hat t_1)\sigma_2(\hat t_2,s_2)
\right]
\\
&+
\one_{\{s_2\in B_2\}}
\left[
f_2(t_2)\sigma_1(t_1,s_1)
-
f_2(\hat t_2)\sigma_1(\hat t_1,s_1)
\right]
\\
&-
\frac{t_1-\hat t_1}{\bar v_1-\ell_1}
\one_{\{\hat t_1,s_1\in B_1\}}
\sigma_2(\hat t_2,s_2)
-
\frac{t_2-\hat t_2}{\bar v_2-\ell_2}
\one_{\{\hat t_2,s_2\in B_2\}}
\sigma_1(\hat t_1,s_1)
\\
&-
\one_{\{s_1\in B_1,s_2\in B_2\}}
\left[
\one_{\{t_1=\bar v_1,t_2=\bar v_2\}}
-
\one_{\{\hat t_1=\bar v_1,\hat t_2=\bar v_2\}}
\right].
\end{aligned}
\]
Since the original DSIC constraint binds, the corresponding single-item IC
constraint also binds for item $1$. By \eqref{eq:reduction-scalar-f}, the terms
associated with item $1$ in the preceding expression simplify to
\[
\one_{\{s_1\in B_1\}}
f_1(t_1)
\bigl[
\sigma_2(t_2,s_2)-\sigma_2(\hat t_2,s_2)
\bigr].
\]
Applying the same argument to item \(2\), and then including 
the difference of the final indicator terms in \(\Delta u_1\), yields
\begin{equation}
\begin{aligned}
&\frac{1}{\varepsilon}
\left[
\Delta u_1(t,s)-\Delta u_1(\hat t,s)
-(t-\hat t)\cdot\Delta x_1(\hat t,s)
\right]
\\
={}&
\one_{\{s_1\in B_1\}}
f_1(t_1)
\bigl[
\sigma_2(t_2,s_2)-\sigma_2(\hat t_2,s_2)
\bigr]
\\
&+
\one_{\{s_2\in B_2\}}
f_2(t_2)
\bigl[
\sigma_1(t_1,s_1)-\sigma_1(\hat t_1,s_1)
\bigr]
\\
&-
\one_{\{s_1\in B_1,s_2\in B_2\}}
\left[
\one_{\{t_1=\bar v_1,t_2=\bar v_2\}}
-
\one_{\{\hat t_1=\bar v_1,\hat t_2=\bar v_2\}}
\right].
\end{aligned}
\label{eq:reduction-two-items-slack}
\end{equation}

The first two terms are nonnegative by
\eqref{eq:reduction-scalar-sigma}. 
The last term can be negative only when
$s_1\in B_1$, $s_2\in B_2$, $t_1=\bar v_1$, $t_2=\bar v_2$,
and \((\hat t_1,\hat t_2)\neq(\bar v_1,\bar v_2)\).

Because the original DSIC constraint binds,
the corresponding one-dimensional IC constraint binds in each coordinate. 
Since \(t_j=\bar v_j\) for \(j=1,2\) and $s_j \in B_j$ for \(j=1,2\), 
\eqref{eq:reduction-scalar-top-reports} implies
\[
\hat t_j\in B_j\cup\{\bar v_j\}
\]
for $j=1,2$. 
As \((\hat t_1,\hat t_2)\neq(\bar v_1,\bar v_2)\), at least one of these reported coordinates must lie in its corresponding block.

Suppose first that \(\hat t_1\in B_1\). Since \(s_1\in B_1\), \(t_1=\bar v_1\), and \(\hat t_1\in B_1\), \eqref{eq:reduction-scalar-block-sigma} gives
\[
\sigma_1(t_1,s_1)-\sigma_1(\hat t_1,s_1)=1.
\]
Moreover, \(t_2=\bar v_2\) implies \(f_2(t_2)=1\). 
Hence the second term in \eqref{eq:reduction-two-items-slack} is \(1\), 
so it cancels the possible \(-1\) from the final term.
If instead \(\hat t_2\in B_2\), then \eqref{eq:reduction-scalar-block-sigma} gives
\[
\sigma_2(t_2,s_2)-\sigma_2(\hat t_2,s_2)=1,
\]
while \(t_1=\bar v_1\) implies \(f_1(t_1)=1\). Hence the first term in \eqref{eq:reduction-two-items-slack} is \(1\), which cancels the possible \(-1\) from the final term.

Thus, whenever the final term is negative, it is offset by at least one of the first two nonnegative terms. In all other cases the final term is itself nonnegative. Therefore the slack in \eqref{eq:reduction-two-items-slack} is always nonnegative, and every originally binding DSIC constraint remains satisfied under the perturbation.

Each originally strict DSIC slack changes linearly with
$\varepsilon$. Since there are finitely many, one sufficiently small
$\varepsilon>0$ preserves them all.

A negative utility change can occur only at a profile where the buyer's
original utility is strictly positive. If a summand in
\eqref{eq:reduction-two-items-utility} is negative for item \(j\), then
\(s_j\in B_j\) and \(f_j(t_j)>0\), so the original utility from that item is
\(q_j(\ell_j,s_j)(t_j-\ell_j)>0\). If the final indicator term is negative,
the buyer values both affected items at their highest values and the other
buyer reports a value in each corresponding block, again giving strictly positive
original utility. Since the type space is finite and \(\Delta u_1\) is linear
in \(\varepsilon\), IR is preserved for sufficiently small \(\varepsilon>0\).

To verify nonnegative payments, note that whenever the perturbation changes the allocation of item \(j\), we have \(t_j,s_j\in B_j\). The buyer's original canonical payment from that item is then \(q_j(\ell_j,s_j)\ell_j>0\). Likewise, whenever the utility perturbation is nonzero, at least one affected item generates a strictly positive original payment. Since the change in payment is linear in \(\varepsilon\), all payments remain nonnegative for sufficiently small \(\varepsilon>0\).

It remains to compute expected revenue. 
Conditional on \(E\), buyer symmetry gives
\[
\mathbb E[\sigma_j(t_j,s_j)\mid E]=0.
\]
By antisymmetry, \(\Delta x_{21}(t,s)=-\Delta x_{11}(t,s)\).
Since $E$ is independent of $t,s$, independence across items gives
\[
\begin{aligned}
&\mathbb E\left[
t_1\Delta x_{11}(t,s)+s_1\Delta x_{21}(t,s)
\mid E\right]\\
&=
\frac{\varepsilon}{\bar v_1-\ell_1}
\mathbb E\left[(t_1-s_1)\one_{\{t_1,s_1\in B_1\}}\right]
\mathbb E\left[\sigma_2(t_2,s_2)\right]
=
0.
\end{aligned}
\]

Similarly, for item \(2\), antisymmetry implies
\[
\Delta x_{22}(t,s)=-\Delta x_{12}(t,s).
\]
Hence,
\[
\begin{aligned}
&\mathbb E\left[
t_2\Delta x_{12}(t,s)+s_2\Delta x_{22}(t,s)
\mid E\right]\\
&=
\frac{\varepsilon}{\bar v_2-\ell_2}
\mathbb E\left[(t_2-s_2)\one_{\{t_2,s_2\in B_2\}}\right]
\mathbb E\left[\sigma_1(t_1,s_1)\right]
=
0.
\end{aligned}
\]

The two summands in \eqref{eq:reduction-two-items-utility} also have
expectation zero. 
The final indicator term, however, lowers the sum of buyers' expected utilities by
\begin{equation}
2\varepsilon
\mathrm{Pr}(E)
\mathrm{Pr}(s_1\in B_1)
\mathrm{Pr}(s_2\in B_2)
p_1(\bar v_1)p_2(\bar v_2)
>0.
\label{eq:reduction-two-items-gain}
\end{equation}
Expected allocated value is unchanged, so expected revenue strictly
increases by \eqref{eq:reduction-two-items-gain}, contradicting optimality of separate selling.

Hence top selling cannot be suboptimal for two or more items.

\medskip
\noindent
\emph{Case 2: top selling is suboptimal for exactly one item.}

Relabel the item for which top selling is suboptimal as item \(1\).
By the standing support assumption,
there exists another item \(2\) satisfying
\begin{equation}
v_{2,K_2-1}>0.
\label{eq:reduction-second-positive-value}
\end{equation}
Since top selling is suboptimal only for item \(1\), the top-selling auction is
optimal for item \(2\).
Because the revenue from top selling is 
independent of how ties among buyers reporting 
\(\bar v_2\) are broken, 
we use uniform tie-breaking for item \(2\).

Use the Myerson auction on item \(1\), and the symmetric
top-selling auction on item \(2\).
For the top-selling auction on item \(2\),
\[
\sigma_2(v,w)
=
\one_{\{v=\bar v_2\}}
-
\one_{\{w=\bar v_2\}}.
\]
Let \(E\) again be the event that buyers
\(3,\ldots,n\) report their lowest support values on these two items.

Write \(t \in \mathcal V\) and \(s \in \mathcal V\) 
for the types of buyers \(1\) and \(2\).
For buyer \(1\), define on \(E\)
\begin{equation}
\Delta x_{11}(t,s)
=
\frac{\varepsilon}{\bar v_1-\ell_1}
\one_{\{t_1,s_1\in B_1\}}
\sigma_2(t_2,s_2),
\label{eq:reduction-one-item-allocation}
\end{equation}
\begin{equation}
\frac{\Delta u_1(t,s)}{\varepsilon}
=
\one_{\{s_1\in B_1\}}
f_1(t_1)
\sigma_2(t_2,s_2),
\label{eq:reduction-one-item-utility}
\end{equation}
and add the item-\(2\) lottery
\begin{equation}
\Delta x_{12}(t,s)
=
\frac{\varepsilon}
{\bar v_2-v_{2,K_2-1}}
\one_{\{
t_1=\bar v_1,\,
s_1\in B_1,\,
t_2=v_{2,K_2-1},\,
s_2<\bar v_2
\}}.
\label{eq:reduction-second-item-lottery}
\end{equation}
Define buyer \(2\)'s changes by exchanging \(t\) and \(s\) in these
formulas, set all other changes to zero, and reconstruct payments
from allocations and utilities as in Case 1.

The perturbation in \eqref{eq:reduction-one-item-allocation} 
only transfers allocation probability 
for item \(1\) between buyers \(1\) and \(2\) 
at profiles where they are tied in the original auction, 
so the total allocation of item \(1\) is unchanged. 
The lottery in \eqref{eq:reduction-second-item-lottery} 
is added only when item \(2\) is unallocated under top selling. 
Moreover, buyers \(1\) and \(2\) cannot both satisfy 
the indicator in \eqref{eq:reduction-second-item-lottery} 
at the same profile. Hence, for sufficiently small 
\(\varepsilon>0\), the perturbed allocation remains feasible.

Fix a DSIC constraint for buyer \(1\) that binds under the original
product mechanism, with true type \(t\) and report \(\hat t\).
Its single-item IC constraints bind in both affected coordinates.
Ignoring the item-\(2\) lottery, \eqref{eq:reduction-scalar-f}
gives the change in its slack, divided by \(\varepsilon\):
\begin{equation}
\one_{\{s_1\in B_1\}}
f_1(t_1)
\left[
\sigma_2(t_2,s_2)-\sigma_2(\hat t_2,s_2)
\right]
\ge0.
\label{eq:reduction-one-item-slack}
\end{equation}
To verify the inequality, consider the binding one-dimensional IC constraint
for item \(2\), with true value \(t_2\), report \(\hat t_2\), and the fixed
opponents' profile on \(E\). Under top selling, a value below \(\bar v_2\)
gets zero utility from truthful reporting. Reporting \(\bar v_2\) instead
wins with positive probability and requires payment \(\bar v_2\) conditional
on winning, so it gives strictly negative utility. Hence
\(\hat t_2=\bar v_2\) in a binding constraint implies
\(t_2=\bar v_2\). The displayed formula for \(\sigma_2\) therefore gives
\[
\sigma_2(t_2,s_2)-\sigma_2(\hat t_2,s_2)
=
\one_{\{t_2=\bar v_2\}}-\one_{\{\hat t_2=\bar v_2\}}
\ge0.
\]
Since \(f_1(t_1)\ge0\), \eqref{eq:reduction-one-item-slack} follows.

Suppose $\Delta x_{12}(\hat t,s)>0$. 
By the definition of the item-\(2\) lottery in 
\eqref{eq:reduction-second-item-lottery}, this can occur only if
\[
\begin{aligned}
\hat t_1&=\bar v_1,&
\hat t_2&=v_{2,K_2-1},\\
s_1&\in B_1,&
s_2&<\bar v_2.
\end{aligned}
\]
Binding in coordinate \(1\), together with \eqref{eq:reduction-scalar-top-reports}, implies
\[
t_1=\bar v_1.
\]
Because the lottery leaves utility unchanged, its contribution to the
right-hand side of the DSIC constraint, divided by \(\varepsilon\), is
\begin{equation}
\frac{t_2-v_{2,K_2-1}}
{\bar v_2-v_{2,K_2-1}}.
\label{eq:reduction-lottery-report-effect}
\end{equation}
If \(t_2\le v_{2,K_2-1}\), this is nonpositive. 

The only remaining case in which the lottery can
reduce the slack of a binding DSIC constraint is
\(t_2=\bar v_2\). In this case,
\eqref{eq:reduction-lottery-report-effect} equals one. Since
\(s_2<\bar v_2\) and
\(\hat t_2=v_{2,K_2-1}<\bar v_2\),
\[
\sigma_2(t_2,s_2)-\sigma_2(\hat t_2,s_2)=1.
\]
Together with \(f_1(t_1)=1\),
\eqref{eq:reduction-one-item-slack} therefore increases the slack
by the same amount. Hence every originally binding DSIC constraint
remains satisfied.
Finiteness and linearity in $\varepsilon$ preserve all originally
strict DSIC constraints for sufficiently small $\varepsilon>0$.

A negative change in \eqref{eq:reduction-one-item-utility}
requires $s_1\in B_1$ and $f_1(t_1)>0$, so the original item-\(1\)
utility is positive. Every nonzero allocation or utility change for
buyer \(1\) requires $s_1\in B_1$ and
$t_1\in B_1\cup\{\bar v_1\}$. The original item-\(1\) canonical
payment is then at least $q_1(\ell_1,s_1)\ell_1>0$.
Finiteness and linearity in $\varepsilon$ preserve IR and
nonnegative payments for sufficiently small $\varepsilon>0$;
buyer \(2\) is symmetric and buyers \(3,\ldots,n\) are unchanged.

By buyer symmetry and item independence, both
$\mathbb E[t_1\Delta x_{11}+s_1\Delta x_{21}]$ and
$\mathbb E[\Delta u_1+\Delta u_2]$ vanish. The
lottery in \eqref{eq:reduction-second-item-lottery} leaves utility
unchanged and yields the strictly positive expected revenue gain
\begin{equation}
\frac{
2\varepsilon v_{2,K_2-1}
}{
\bar v_2-v_{2,K_2-1}
}
\mathrm{Pr}(E)
p_1(\bar v_1)
\left(\sum_{v\in B_1}p_1(v)\right)
p_2(v_{2,K_2-1})
\bigl[1-p_2(\bar v_2)\bigr]
>0,
\label{eq:reduction-second-item-gain}
\end{equation}
where strict positivity uses \eqref{eq:reduction-second-positive-value}
and the positive probability of every support value.

The gain in \eqref{eq:reduction-second-item-gain} again contradicts optimality of separate selling.

We conclude that, for every \(j\in[m]\), the single-item top-selling auction
is optimal, so
\[
\Rev_n(F_j)
=
\bar v_j
\left[
1-(1-p_j(\bar v_j))^n
\right].
\]
Summing over items,
\[
\Rev_n(\mathbf F)
=
\SREV_n(\mathbf F)
=
\sum_{j\in[m]}
\bar v_j
\left[
1-(1-p_j(\bar v_j))^n
\right].
\]
Thus top selling is optimal for $n$ buyers, proving (iii).

\emph{(iii)$\Rightarrow$(iv).}
\begin{claim}\label{clm:reduction-many-to-one}
If top selling is optimal for $n\ge2$ buyers, it is optimal for one buyer.
\end{claim}

\begin{proof}[Proof of Claim~\ref{clm:reduction-many-to-one}]
Argue by contraposition.
Suppose there exists a one-buyer IC and IR mechanism \((q,\pi)\)
whose expected revenue is strictly greater than that of one-buyer
top selling.
Let $E$ be the event that buyers $2,\ldots,n$ report below the top value
on every item $j\in[m]$.
This event has positive probability
and is independent of buyer 1's type.
On $E$, run $(q,\pi)$ for buyer 1 and give every other buyer nothing.
On $E^c$, run the top-selling mechanism. Because \(E\) depends only on the opponents'
reports, buyer 1 cannot affect which mechanism is used by changing her
own report.
Thus, for every fixed opponents' profile, she faces an IC and IR menu.
Every other buyer obtains zero utility from truthful reporting regardless
of which mechanism is used.
After any deviation by a buyer other than buyer 1, that buyer either
receives nothing when \((q,\pi)\) is used or, under top selling,
receives only items priced at their highest support values.
Since the true value of any buyer other than buyer 1
for each such item is no greater than its highest support value,
any such deviation gives that buyer utility at most zero.
Hence, no deviation that changes which mechanism is used can
be profitable.

The constructed mechanism is therefore DSIC and IR.
For every item \(j\in[m]\), only buyer \(1\) can report
\(\bar v_j\) on \(E\). Hence, conditional on \(E\), top selling generates
expected revenue
\[
\sum_{j\in[m]}\bar v_j p_j(\bar v_j),
\]
which is exactly the expected revenue from one-buyer top selling.
Since \(E\) has positive probability, the strict revenue improvement
on \(E\), together with equality with top selling on \(E^c\),
implies a strict improvement in unconditional expected revenue.
This contradicts (iii), and therefore (iv) follows.
\end{proof}

\emph{(iv)$\Rightarrow$(i).}
Let $(x,\tau)\in\mathcal M^B_{n,m}$. For each buyer $i$, define
the interim allocation and payment at report $t\in\mathcal V$ by
\[
\begin{aligned}
q_i(t)&:=\mathbb E_{v_{-i}}[x_i(t,v_{-i})],\\
\pi_i(t)&:=\mathbb E_{v_{-i}}[\tau_i(t,v_{-i})].
\end{aligned}
\]
The menu $(q_i,\pi_i)$ is IC by BIC and IR by BIR. Its payments are
nonnegative because $\tau_i(v)\ge0$ at every profile. By
Lemma~\ref{lem:top-value-payment-bound} and independence of buyer types,
\[
\mathbb E_v\left[\sum_{i\in[n]}\tau_i(v)\right]
\le
\mathbb E_v\left[
\sum_{j\in[m]}\bar v_j
\sum_{i\in[n]}\one_{\{v_{ij}=\bar v_j\}}x_{ij}(v)
\right].
\]
For every item $j\in[m]$ and profile $v$, feasibility implies
\[
\sum_{i\in[n]}
\one_{\{v_{ij}=\bar v_j\}}x_{ij}(v)
\le
\one_{\{\exists i\in[n]:v_{ij}=\bar v_j\}}.
\]
Consequently, the revenue of $(x,\tau)$ is at most
\[
\sum_{j\in[m]}\bar v_j
\mathrm{Pr}(\exists i\in[n]:v_{ij}=\bar v_j),
\]
the expected revenue from top selling. Separate selling earns at least
as much as top selling and is BIC and BIR, proving (i).
\end{proof}

\subsection{One-buyer dual representation}
\label{app:one-buyer-dual}

Let $\lambda$ be a nonnegative flow satisfying
\eqref{eq:flow:f1}; top preservation and \eqref{eq:flow:f2} are not
imposed. For each $t\in\mathcal V$, define
\[
b_\lambda(t):=\mathrm{Pr}(t)+
\sum_{\hat t\in\mathcal V}
[\lambda(t,\hat t)-\lambda(\hat t,t)].
\]
Thus $b_\lambda(t)\ge0$ by \eqref{eq:flow:f1}. For each
$t\in\mathcal V$ and $j\in[m]$, define
\[
\varphi_j^\lambda(t):=
t_j-\frac{\sum_{\hat t\in\mathcal V}
\lambda(t,\hat t)(\hat t_j-t_j)}{\mathrm{Pr}(t)}.
\]

\begin{lemma}[One-buyer dual representation]
\label{lem:one-buyer-dual}
The optimal one-buyer revenue satisfies
\begin{equation}
\Rev_1(\mathbf F)=\min_{\lambda\ge0}
\mathbb E_t\left[\sum_{j\in[m]}(\varphi_j^\lambda(t))_+\right],
\label{eq:one-buyer-dual}
\end{equation}
where the minimum is over flows satisfying \eqref{eq:flow:f1} and is
attained.
\end{lemma}

\begin{proof}
First, signed payments cannot increase optimal revenue. For an IC and
IR menu $(q,\pi)$, write $u(t)=t\cdot q(t)-\pi(t)$. IC in both
directions between $t$ and $\underline v$ gives
\[
0\le u(t)-u(\underline v)
\le(t-\underline v)\cdot q(t).
\]
Replace each payment $\pi(t)$ by $\pi(t)+u(\underline v)$.
The new utility is $u(t)-u(\underline v)\ge0$, so IC and IR hold.
The new payment is at least $\underline v\cdot q(t)\ge0$.
Because $u(\underline v)\ge0$, expected revenue weakly increases.

We may therefore solve the one-buyer problem over allocations
$q_j(t)\in[0,1]$ and utilities $u(t)\ge0$, allowing signed
payments in the linear program. The objective is
\[
\sum_{t\in\mathcal V}\mathrm{Pr}(t)
\left[\sum_{j\in[m]}t_jq_j(t)-u(t)\right].
\]
For every true type $t$ and report $\hat t$, the IC constraint is
\[
u(\hat t)-u(t)+(t-\hat t)\cdot q(\hat t)\le0.
\]
Assign it a nonnegative multiplier $\lambda(\hat t,t)$ and subtract
the weighted constraints from the objective. The coefficient on
$u(t)$ is $-b_\lambda(t)$. Since $u(t)$ has no upper bound, the
Lagrangian has a finite maximum only if $b_\lambda(t)\ge0$ for every
$t$, which is \eqref{eq:flow:f1}. The coefficient on $q_j(t)$ is
$\mathrm{Pr}(t)\varphi_j^\lambda(t)$. Maximizing over
$q_j(t)\in[0,1]$ therefore yields the right-hand objective in
\eqref{eq:one-buyer-dual}. The primal is feasible and has revenue
bounded by expected total value. Finite-dimensional linear-programming
strong duality gives equality and attainment of the dual minimum.
\end{proof}

\subsection{Proof of the flow characterization}
\label{app:flow-characterization-proof}

\begin{proof}[Proof of Proposition~\ref{prop:flow-characterization}]
Let $\lambda$ be any nonnegative flow satisfying \eqref{eq:flow:f1}.
If $t_j=\bar v_j$, then
\[
\varphi_j^\lambda(t)=\bar v_j+
\frac{\sum_{\hat t\in\mathcal V}\lambda(t,\hat t)(\bar v_j-\hat t_j)}
{\mathrm{Pr}(t)}\ge\bar v_j.
\]
Consequently, for every $t\in\mathcal V$,
\[
\sum_{j\in[m]}(\varphi_j^\lambda(t))_+
\ge\sum_{\substack{j\in[m]\\t_j=\bar v_j}}\bar v_j.
\]
Taking expectations of the preceding inequality yields
\[
\mathbb E_t\left[\sum_{j\in[m]}(\varphi_j^\lambda(t))_+\right]
\ge
\sum_{j\in[m]}\bar v_j\,\mathrm{Pr}(t_j=\bar v_j)
=
\sum_{j\in[m]}\bar v_jp_j(\bar v_j).
\]
Thus, the one-buyer dual objective is bounded below by the revenue from top selling.
Since every type
$t\in\mathcal V$ has positive probability, equality holds
precisely when $\varphi_j^\lambda(t)=\bar v_j$ for every $t\in\mathcal V$
and $j\in[m]$ with $t_j=\bar v_j$, and $\varphi_j^\lambda(t)\le0$
for every $t\in\mathcal V$ and $j\in[m]$ with $t_j<\bar v_j$.
The first condition is equivalent to top preservation, since the
displayed numerator is a sum of nonnegative terms. The second is
\eqref{eq:flow:f2}, and \eqref{eq:flow:f1} is already imposed.
By Lemma~\ref{lem:one-buyer-dual}, the dual minimum is attained and equals
the optimal one-buyer revenue. Therefore, top selling is optimal if and only
if some flow attains the lower bound above. By the equality conditions, this
holds precisely when the flow is top-preserving and satisfies
\eqref{eq:flow:f2}, in addition to \eqref{eq:flow:f1}.

\end{proof}

\subsection{Proofs of the flow conditions}
\label{app:flow-proofs}

The following lemma constructs flows between ordered values. It will be
applied separately to each item's support in the proofs of
Propositions~\ref{prop:upward} and~\ref{prop:boundary-flow}.

Think of $a_\ell$ as the additional amount of flow that $z_\ell$ may
receive beyond what it sends, and $d_k$ as the required value-weighted
amount sent upward from $z_k$ in the construction below.
Condition (ii) in the lemma below says that, for every $k\in[K-1]$,
the capacities at values above $z_k$, weighted by their distances from $z_k$, cover all
requirements $d_k,\ldots,d_{K-1}$.

\begin{lemma}\label{lem:one-dimensional-capacity}
Let $z_1<\cdots<z_K$ be real numbers, where $K\ge1$, and let
$a\in\mathbb R_+^K$ and $d\in\mathbb R_+^{K-1}$. The following are
equivalent.
\begin{enumerate}[label=(\roman*)]
\item There exists a matrix $f=(f_{k\ell})_{k,\ell\in[K]}\in
\mathbb R_+^{K\times K}$ such that, for every $\ell\in[K]$,
\begin{equation}
\begin{aligned}
f_{K\ell}&=0,\\
\sum_{r\in[K]}f_{r\ell}&\le a_\ell+\sum_{r\in[K]}f_{\ell r},
\end{aligned}
\label{eq:one-dimensional-sums}
\end{equation}
and, for every $k\in[K-1]$,
\begin{equation}
\sum_{\ell\in[K]}f_{k\ell}(z_\ell-z_k)\ge d_k.
\label{eq:one-dimensional-weighted-sum}
\end{equation}
\item For every $k\in[K-1]$,
\begin{equation}
\sum_{\ell=k+1}^K a_\ell(z_\ell-z_k)\ge\sum_{r=k}^{K-1}d_r.
\label{eq:one-dimensional-capacity}
\end{equation}
\end{enumerate}
When these conditions hold, the matrix in (i) can be chosen to satisfy
$f_{k\ell}=0$
for every $k,\ell\in[K]$ with $\ell\le k$, and
\begin{equation}
\sum_{\ell=k+1}^{K}f_{k\ell}(z_\ell-z_k)=d_k
\label{eq:one-dimensional-exact}
\end{equation}
for every $k\in[K-1]$.
\end{lemma}
\begin{proof}
We may take $f_{kk}=0$ for every $k\in[K]$, since this leaves
\eqref{eq:one-dimensional-sums} and
\eqref{eq:one-dimensional-weighted-sum} unchanged. 
To prove that (i) implies (ii), fix $k\in[K-1]$. 
Multiplying the inequality in \eqref{eq:one-dimensional-sums} 
by \((z_\ell-z_k)_+\) and summing over \(\ell\in[K]\) gives
\[
\begin{aligned}
\sum_{\ell=k+1}^{K}a_\ell(z_\ell-z_k)
&\ge
\sum_{\ell\in[K]}(z_\ell-z_k)_+
\left[
\sum_{r\in[K]}f_{r\ell}
-
\sum_{r\in[K]}f_{\ell r}
\right] \\
&=
\sum_{r\in[K]}\sum_{\ell\in[K]}f_{r\ell}
\left[
(z_\ell-z_k)_+
-
(z_r-z_k)_+
\right].
\end{aligned}
\]
For every $r,\ell\in[K]$, apply the subgradient inequality to
$g_k(u):=(u-z_k)_+$ at $z_r$, using
\(\one_{\{r\ge k\}}\in\partial g_k(z_r)\):
\[
(z_\ell-z_k)_+-(z_r-z_k)_+
\ge\one_{\{r\ge k\}}(z_\ell-z_r).
\]
Since $f_{K\ell}=0$ for
every $\ell\in[K]$, applying
\eqref{eq:one-dimensional-weighted-sum} for each $r$ with $k\le r<K$ yields
\[
\begin{aligned}
\sum_{\ell=k+1}^{K}a_\ell(z_\ell-z_k)
&\ge\sum_{r\in[K-1]}\sum_{\ell\in[K]}f_{r\ell}
[(z_\ell-z_k)_+-(z_r-z_k)_+]\\
&\ge\sum_{r=k}^{K-1}\sum_{\ell\in[K]}f_{r\ell}(z_\ell-z_r)
\ge\sum_{r=k}^{K-1}d_r.
\end{aligned}
\]

To prove that (ii) implies (i), we construct \(f \in \mathbb R_+^{K \times K}\) 
row by row. Set 
$f_{k\ell}=0$
whenever $\ell\le k$.
Thus, it remains to choose \(f_{k\ell}\) for \(\ell>k\). 
We do so in the order 
$k=K-1,K-2,\ldots,1$.

At step \(k\), rows \(k+1,\ldots,K-1\) have already been chosen, 
and our goal is to choose \(f_{k\ell}\) for \(\ell>k\), so that 
$\sum_{\ell=k+1}^K f_{k\ell}(z_\ell-z_k)=d_k$,
while preserving the inequalities in \eqref{eq:one-dimensional-sums}.

Fix \(\ell>k\). Given the rows \(k+1,\ldots,K-1\) already constructed, the constraint in \eqref{eq:one-dimensional-sums} for column \(\ell\) is
\[
\sum_{r=k+1}^{\ell-1}f_{r\ell}
+
f_{k\ell}
\le
a_\ell+\sum_{r=\ell+1}^{K}f_{\ell r}.
\]
Therefore, \(f_{k\ell}\) may be chosen up to
\[
c_{k\ell}
:=
a_\ell+\sum_{r=\ell+1}^{K}f_{\ell r}
-\sum_{r=k+1}^{\ell-1}f_{r\ell}.
\]
By induction, \(c_{k\ell}\ge0\) for every \(\ell>k\).
We claim that
\[
\sum_{\ell=k+1}^K c_{k\ell}(z_\ell-z_k)\ge d_k.
\]
Indeed,
\[
\begin{aligned}
\sum_{\ell=k+1}^K c_{k\ell}(z_\ell-z_k)
&=
\sum_{\ell=k+1}^K
(z_\ell-z_k)
\left[
a_\ell
+\sum_{r=\ell+1}^{K}f_{\ell r}
-\sum_{r=k+1}^{\ell-1}f_{r\ell}
\right]\\
&=
\sum_{\ell=k+1}^K a_\ell(z_\ell-z_k)
-
\sum_{r=k+1}^{K-1}
\sum_{\ell=r+1}^K
f_{r\ell}(z_\ell-z_r).
\end{aligned}
\]
For every previously constructed row \(r>k\), the induction hypothesis gives
\[
\sum_{\ell=r+1}^K
f_{r\ell}(z_\ell-z_r)
=
d_r.
\]
Hence
\[
\sum_{\ell=k+1}^K c_{k\ell}(z_\ell-z_k)
=
\sum_{\ell=k+1}^K a_\ell(z_\ell-z_k)
-
\sum_{r=k+1}^{K-1}d_r.
\]
By condition \eqref{eq:one-dimensional-capacity},
$\sum_{\ell=k+1}^K a_\ell(z_\ell-z_k)\ge\sum_{r=k}^{K-1}d_r$
and therefore 
$\sum_{\ell=k+1}^K c_{k\ell}(z_\ell-z_k)\ge d_k$.

If \(d_k=0\), set \(f_{k\ell}=0\) for every \(\ell>k\). 
If \(d_k>0\), then $\sum_{\ell=k+1}^K c_{k\ell}(z_\ell-z_k)\ge d_k>0$.
This means 
$\frac{d_k}{\sum_{\ell=k+1}^K c_{k\ell}(z_\ell-z_k)}\le1$.
Set
\[
f_{k\ell}:=\frac{d_k}{\sum_{\ell=k+1}^K c_{k\ell}(z_\ell-z_k)} c_{k\ell}
\]
for $\ell>k$.
Then
$0\le f_{k\ell}\le c_{k\ell}$,
so all column constraints remain satisfied, and
\[
\sum_{\ell=k+1}^K
f_{k\ell}(z_\ell-z_k)
=
\frac{d_k}{\sum_{\ell=k+1}^K c_{k\ell}(z_\ell-z_k)} \sum_{\ell=k+1}^K c_{k\ell}(z_\ell-z_k)
=
d_k.
\]
Moreover, after this choice the unused slack in column \(\ell>k\) is
$c_{k\ell}-f_{k\ell}\ge0$.
Thus the same argument can be repeated at step \(k-1\).

Starting from \(k=K-1\) and continuing down to \(k=1\) 
yields a nonnegative matrix \(f\) satisfying 
\eqref{eq:one-dimensional-sums} and
$\sum_{\ell=k+1}^K f_{k\ell}(z_\ell-z_k)=d_k$
for every \(k\in[K-1]\). Hence \(f\) satisfies (i), and it also satisfies \eqref{eq:one-dimensional-exact}.
\end{proof}

\begin{proof}[Proof of Proposition~\ref{prop:upward}]
For every $j\in[m]$ and $k\in[K_j-1]$, write
\[
\begin{aligned}
N_{j,k}&:=\sum_{\ell=k}^{K_j-1}p_j(v_{j,\ell})v_{j,\ell},\\
D_{j,k}&:=\sum_{\ell=k+1}^{K_j}p_j(v_{j,\ell})(v_{j,\ell}-v_{j,k}).
\end{aligned}
\]
For convenience in this and the later proofs, define for each $j\in[m]$
\[
\rho_j:=\max_{1\le k<K_j}\rho_{j,k}
=\max_{1\le k<K_j}\frac{N_{j,k}}{D_{j,k}}.
\]

Suppose \eqref{eq:slice-tail} holds, equivalently,
$\sum_{j\in[m]}\rho_j\le1$.
We construct an upward flow for each
coordinate $j \in [m]$. 
For item $j$, apply Lemma~\ref{lem:one-dimensional-capacity}
with $z_\ell=v_{j,\ell}$ and $a_\ell=\rho_jp_j(v_{j,\ell})$ for $\ell\in[K_j]$,
and $d_k=p_j(v_{j,k})v_{j,k}$ for $k\in[K_j-1]$.
With these definitions, condition~\eqref{eq:one-dimensional-capacity} becomes
$\rho_jD_{j,k}\ge N_{j,k}$ for every $k\in[K_j-1]$.
These inequalities hold by definition of $\rho_j$. For each $j\in[m]$,
the last assertion of the lemma gives a matrix
$f^j=(f^j_{k\ell})_{k,\ell\in[K_j]}$.
For each $j\in[m]$ and $t,\hat t\in\mathcal V$ with $t_j=v_{j,k}$, set
\[
\lambda^j(t,\hat t):=\frac{\mathrm{Pr}(t)}{p_j(v_{j,k})}f^j_{k\ell}
\]
when $\hat t_j=v_{j,\ell}>t_j$ and $\hat t_h=t_h$ for every
$h\in[m]\setminus\{j\}$, and set $\lambda^j(t,\hat t)=0$ for all other
pairs. Define the combined flow by
\[
\lambda(t,\hat t):=\sum_{j\in[m]}\lambda^j(t,\hat t)
\]
for every $t,\hat t\in\mathcal V$.
For every $t\in\mathcal V$ and $j\in[m]$, write $t_j=v_{j,k}$.
Since values are independent across items,
\[
\frac{\mathrm{Pr}(t)}{p_j(t_j)}
=
\prod_{h\ne j}p_h(t_h).
\]
Hence this ratio is the same for all types that differ only in coordinate \(j\). In particular, if \(t_j=v_{j,k}\) and \(\hat t_j=v_{j,\ell}\) while \(\hat t_h=t_h\) for all \(h\ne j\), then
\[
\frac{\mathrm{Pr}(\hat t)}{p_j(v_{j,\ell})}
=
\frac{\mathrm{Pr}(t)}{p_j(v_{j,k})}.
\]
Therefore,
\[
\sum_{\hat t\in\mathcal V}
[\lambda^j(\hat t,t)-\lambda^j(t,\hat t)]
=
\frac{\mathrm{Pr}(t)}{p_j(v_{j,k})}
\sum_{\ell\in[K_j]}
(f^j_{\ell k}-f^j_{k\ell}).
\]
The lemma gives
\[
\begin{aligned}
\sum_{\hat t\in\mathcal V}
[\lambda^j(\hat t,t)-\lambda^j(t,\hat t)]
\le\rho_j\mathrm{Pr}(t).
\end{aligned}
\]
Summing these inequalities over $j\in[m]$ and using
$\sum_{j\in[m]}\rho_j\le1$ proves \eqref{eq:flow:f1}.
For every $h\in[m]\setminus\{j\}$, a positive value of
$\lambda^h(t,\hat t)$ requires $\hat t_j=t_j$. Hence, for
$t_j<\bar v_j$, the equality in the lemma gives
\[
\begin{aligned}
\sum_{\hat t\in\mathcal V}\lambda(t,\hat t)(\hat t_j-t_j)
&=\sum_{\hat t\in\mathcal V}\lambda^j(t,\hat t)(\hat t_j-t_j)\\
&=\frac{\mathrm{Pr}(t)}{p_j(v_{j,k})}
\sum_{\ell\in[K_j]}f^j_{k\ell}(v_{j,\ell}-v_{j,k})\\
&=\mathrm{Pr}(t)t_j.
\end{aligned}
\]
This proves \eqref{eq:flow:f2}, and $\lambda$ is upward by construction.

Conversely, let a feasible upward flow $\lambda$ be given. Fix an arbitrary
$k_j\in[K_j-1]$ for each $j\in[m]$.
For every $t\in\mathcal V$, define
\[
H(t):=\prod_{j=1}^m (t_j-v_{j,k_j})_+.
\]
Whenever $\lambda(t,\hat t)>0$, every factor in $H(\hat t)$ is at least
the corresponding factor in $H(t)$.
Write
$\Delta_j:=(\hat t_j-v_{j,k_j})_+-(t_j-v_{j,k_j})_+ \ge 0$ for 
each \(j\in[m]\).
Therefore
\[
H(\hat t)-H(t)
=
\sum_{\varnothing\ne S\subseteq[m]}
\left(\prod_{j\in S}\Delta_j\right)
\left(\prod_{h\notin S}(t_h-v_{h,k_h})_+\right).
\]
Every term on the right-hand side is nonnegative. 
Keeping only the terms corresponding to singleton sets \(S=\{j\}\) therefore yields
\[
H(\hat t)-H(t)
\ge
\sum_{j\in[m]}
\Delta_j
\prod_{h\in[m]\setminus\{j\}}(t_h-v_{h,k_h})_+.
\]

If \(v_{j,k_j}\le t_j<\bar v_j\), then upwardness implies
\(\hat t_j\ge t_j\ge v_{j,k_j}\), and hence
\[
\Delta_j
=
(\hat t_j-v_{j,k_j})-(t_j-v_{j,k_j})
=
\hat t_j-t_j.
\]
If \(t_j=\bar v_j\), upwardness implies
\(\hat t_j=t_j=\bar v_j\), so \(\Delta_j=0\).
Finally, if \(t_j<v_{j,k_j}\), the corresponding term is nonnegative and may be discarded.
Therefore,
\[
H(\hat t)-H(t)\ge
\sum_{\substack{j\in[m]\\v_{j,k_j}\le t_j<\bar v_j}}
(\hat t_j-t_j)
\prod_{h\in[m]\setminus\{j\}}
(t_h-v_{h,k_h})_+.
\]

We now use this lower bound together with the flow constraints.
Multiply \eqref{eq:flow:f1} at each type \(t\) by \(H(t)\ge0\) and sum over \(t\in\mathcal V\). 
This gives
\[
\begin{aligned}
\sum_{t\in\mathcal V}\mathrm{Pr}(t)H(t)
\ge
\sum_{t\in\mathcal V}
H(t)
\sum_{\hat t\in\mathcal V}
[\lambda(\hat t,t)-\lambda(t,\hat t)]
=
\sum_{t,\hat t\in\mathcal V}
\lambda(t,\hat t)
[H(\hat t)-H(t)],
\end{aligned}
\]
where the equality follows by reindexing the inflow term.
Applying the preceding lower bound on \(H(\hat t)-H(t)\) yields
\[
\begin{aligned}
\sum_{t\in\mathcal V}\mathrm{Pr}(t)H(t)
\ge
\sum_{j\in[m]}
\sum_{\substack{t\in\mathcal V\\
v_{j,k_j}\le t_j<\bar v_j}}
\left[
\sum_{\hat t\in\mathcal V}
\lambda(t,\hat t)(\hat t_j-t_j)
\right]
\prod_{h\ne j}(t_h-v_{h,k_h})_+.
\end{aligned}
\]
For every type appearing in the inner sum, \(t_j<\bar v_j\), so
\eqref{eq:flow:f2} implies
$\sum_{\hat t\in\mathcal V}\lambda(t,\hat t)(\hat t_j-t_j)\ge\mathrm{Pr}(t)t_j$.
Hence
\[
\begin{aligned}
\sum_{t\in\mathcal V}\mathrm{Pr}(t)H(t)
\ge
\sum_{j\in[m]}
\sum_{t\in\mathcal V}
\mathrm{Pr}(t)t_j
\one_{\{v_{j,k_j}\le t_j<\bar v_j\}}
\prod_{h\ne j}(t_h-v_{h,k_h})_+.
\end{aligned}
\]
By independence across items,
$\mathbb E[(t_j-v_{j,k_j})_+]=D_{j,k_j}$
and
$\mathbb E[t_j\one_{\{v_{j,k_j}\le t_j<\bar v_j\}}]=N_{j,k_j}$.
Therefore the preceding inequality becomes
\[
\prod_{j\in[m]}D_{j,k_j}
\ge
\sum_{j\in[m]}
N_{j,k_j}
\prod_{h\in[m]\setminus\{j\}}D_{h,k_h}.
\]
Since \(D_{j,k_j}>0\) for every $j$, dividing both sides by
\(\prod_{j\in[m]}D_{j,k_j}\) gives
\[
\sum_{j\in[m]}
\frac{N_{j,k_j}}{D_{j,k_j}}
\le1.
\]
The indices \(k_j\in[K_j-1]\) were arbitrary. 
Choosing, for each \(j\), an index \(k_j\) attaining
$\rho_j=\max_{1\le k<K_j}\frac{N_{j,k}}{D_{j,k}}$
therefore gives
\[
\sum_{j\in[m]}\rho_j\le1.
\]
Thus the slice-tail condition holds.
\end{proof}

We next prove that TTRB exactly characterizes feasible top-preserving flows
when the type space is restricted to types with at most one item below its
highest value.

Let $\mathcal B$ denote the types with at most one item below its highest value:
\[
\mathcal B:=\{t\in\mathcal V:|\{j\in[m]:t_j<\bar v_j\}|\le1\}.
\]

We can formulate the flow problem on this restricted type space.
A flow $\lambda:\mathcal B\times\mathcal B\to\mathbb R_+$ is feasible
if it satisfies \eqref{eq:flow:f1} and \eqref{eq:flow:f2}, with
$\mathcal V$ replaced by $\mathcal B$.
The probabilities \(\mathrm{Pr}(t)\) retain their original values and are not
renormalized on \(\mathcal B\).
Top preservation has the same definition on \(\mathcal B\).

The restriction to \(\mathcal B\) makes the flow problem essentially
one-dimensional.

\begin{proposition}\label{prop:boundary-flow}
A feasible top-preserving flow on $\mathcal B\times\mathcal B$ exists if
and only if TTRB, \eqref{eq:ttrb}, holds. Whenever TTRB holds, it can be
chosen so that \(\lambda(t,\hat t)>0\) implies
\(\hat t_j\ge t_j\) for every \(j\in[m]\).
\end{proposition}

\begin{proof}[Proof of Proposition~\ref{prop:boundary-flow}]
The proof proceeds by decomposing the flow problem on
$\mathcal B\times\mathcal B$ into a
one-dimensional problem for each item and then combining these problems
at the common top type $\bar v$.
For convenience in this proof, define for each $j\in[m]$
\[
\eta_j:=\max_{1\le k<K_j}\eta_{j,k}.
\]

For each $j\in[m]$ and $k\in[K_j]$, let $t^{j,k}$ denote the type with
$t^{j,k}_j=v_{j,k}$ and $t^{j,k}_h=\bar v_h$ for every
$h\in[m]\setminus\{j\}$. Thus $t^{j,k}\in\mathcal B$,
$t^{j,K_j}=\bar v$, and
\[
\mathrm{Pr}(t^{j,k})
=
\frac{\mathrm{Pr}(\bar v)}{p_j(\bar v_j)}p_j(v_{j,k}).
\]

Given any feasible top-preserving flow $\lambda$ on
$\mathcal B\times\mathcal B$, we may set $\lambda(t,t)=0$ for every
$t\in\mathcal B$, since self-flow affects neither
\eqref{eq:flow:f1} nor \eqref{eq:flow:f2}. 
At $\bar v$, every coordinate is at its top value. Top preservation
therefore permits flow only from $\bar v$ to itself, which we have
already set to zero. Hence no flow leaves $\bar v$.

Now fix
$t^{j,k}\ne\bar v$. All coordinates other than $j$ are already at their
top values. Hence, if $\lambda(t^{j,k},\hat t)>0$, top preservation
requires
$\hat t_h=\bar v_h$
for every $h\in[m]\setminus\{j\}$.
Since \(\hat t_j\in V_j\), we have \(\hat t=t^{j,\ell}\) for some \(\ell\in[K_j]\).

Since the sets $\{t^{j,k}:k\in[K_j]\}$ for distinct items $j$ 
intersect only at $\bar v$, the flow problem can be analyzed separately 
for each $j$, except for the common requirement \eqref{eq:flow:f1} at $\bar v$.

Fix $j\in[m]$. For a given feasible top-preserving flow $\lambda$, set
\[
\omega_j:=\sum_{k=1}^{K_j-1}\lambda(t^{j,k},\bar v).
\]
In the converse construction, $\omega_j\ge0$ instead denotes a
prescribed upper bound on the flow to $\bar v$ from these types.

To apply Lemma~\ref{lem:one-dimensional-capacity}, let
$z_\ell=v_{j,\ell}$,
$d_k=p_j(v_{j,k})v_{j,k}$ 
for $k\in[K_j-1]$, and set
$a_\ell=p_j(v_{j,\ell})$ 
for $\ell<K_j$ and 
$a_{K_j}:=\frac{\omega_jp_j(\bar v_j)}{\mathrm{Pr}(\bar v)}$.
Given a flow $\lambda$ and $j \in [m]$,  define
\[
f_{k\ell}:=\frac{p_j(\bar v_j)}{\mathrm{Pr}(\bar v)}
\lambda(t^{j,k},t^{j,\ell}),
\]
for $k,\ell\in[K_j]$.
Under this rescaling, the flow into $\bar v$ satisfies
\[
\sum_{r\in[K_j]}f_{rK_j}\le a_{K_j},
\]
with equality for the given flow in the necessity direction. Top
preservation also implies
$f_{K_j\ell}=0$ for every $\ell\in[K_j]$.
Moreover, condition \eqref{eq:flow:f1} at each type
$t^{j,\ell}$ with $\ell\in[K_j-1]$, together with condition
\eqref{eq:flow:f2} at each type $t^{j,k}$ with
$k\in[K_j-1]$, is exactly the requirement in part (i) of
Lemma~\ref{lem:one-dimensional-capacity}.

Conversely, suppose that \eqref{eq:one-dimensional-capacity} holds.
By Lemma~\ref{lem:one-dimensional-capacity}, there exists an upward
matrix $f=(f_{k\ell})_{k,\ell\in[K_j]}$ satisfying condition~(i).
Rescaling this matrix back to the original flow units, define
\[
\lambda(t^{j,k},t^{j,\ell})
:=
\frac{\mathrm{Pr}(\bar v)}{p_j(\bar v_j)}f_{k\ell},
\]
for $1\le k<\ell\le K_j$ 
and set all other flow values to zero. 
By construction, this flow is upward. 
Moreover, condition
\eqref{eq:flow:f1} holds at each type $t^{j,\ell}$ with
$\ell\in[K_j-1]$, the total flow received at $\bar v=t^{j,K_j}$
from the types $t^{j,k}$ with $k\in[K_j-1]$ is at most $\omega_j$,
and condition \eqref{eq:flow:f2} holds with equality at each type
$t^{j,k}$ with $k\in[K_j-1]$.

We have therefore shown that a flow on the types
$\{t^{j,k}:k\in[K_j]\}$ satisfying \eqref{eq:flow:f1} at each
$t^{j,\ell}$ with $\ell\in[K_j-1]$, satisfying
\eqref{eq:flow:f2} at each $t^{j,k}$ with $k\in[K_j-1]$, and sending
at most $\omega_j$ units of flow into $\bar v$ exists if and only if
condition \eqref{eq:one-dimensional-capacity} of
Lemma~\ref{lem:one-dimensional-capacity} holds.

Under our choices of $z_\ell$, $a_\ell$, and $d_k$, that condition
requires, for every $k\in[K_j-1]$,
\[
\sum_{\ell=k+1}^{K_j-1}
p_j(v_{j,\ell})(v_{j,\ell}-v_{j,k})
+
\frac{\omega_jp_j(\bar v_j)}{\mathrm{Pr}(\bar v)}
(\bar v_j-v_{j,k})
\ge
\sum_{r=k}^{K_j-1}p_j(v_{j,r})v_{j,r}.
\]
Moving the first sum on the left-hand side to the right gives
\[
\begin{aligned}
\frac{\omega_jp_j(\bar v_j)}{\mathrm{Pr}(\bar v)}
(\bar v_j-v_{j,k})
&\ge
\sum_{r=k}^{K_j-1}p_j(v_{j,r})v_{j,r}
-\sum_{\ell=k+1}^{K_j-1}
p_j(v_{j,\ell})(v_{j,\ell}-v_{j,k})\\
&=
\sum_{\ell=k}^{K_j-1}
p_j(v_{j,\ell})v_{j,k}.
\end{aligned}
\]

Since $p_j(\bar v_j)(\bar v_j-v_{j,k})>0$, the preceding inequality is
equivalent, for each $k\in[K_j-1]$, to
\[
\omega_j
\ge
\mathrm{Pr}(\bar v)
\frac{
\sum_{\ell=k}^{K_j-1}p_j(v_{j,\ell})v_{j,k}
}{
p_j(\bar v_j)(\bar v_j-v_{j,k})
}.
\]
Therefore, these inequalities hold simultaneously for every
$k\in[K_j-1]$ if and only if
\[
\omega_j
\ge
\mathrm{Pr}(\bar v)
\max_{1\le k<K_j}
\frac{
\sum_{\ell=k}^{K_j-1}p_j(v_{j,\ell})v_{j,k}
}{
p_j(\bar v_j)(\bar v_j-v_{j,k})
}=\mathrm{Pr}(\bar v)\eta_j.
\]

It remains to impose feasibility at the type $\bar v$.
For necessity, $\omega_j$ is the actual flow received at $\bar v$
from the types $t^{j,k}$ with $k\in[K_j-1]$.
These types exhaust $\mathcal B\setminus\{\bar v\}$, and no flow leaves
$\bar v$. Hence condition \eqref{eq:flow:f1} at $\bar v$ gives
\[
\sum_{j\in[m]}\omega_j
=\sum_{j\in[m]}\sum_{k=1}^{K_j-1}\lambda(t^{j,k},\bar v)
\le\mathrm{Pr}(\bar v).
\]

Conversely, suppose that there exist nonnegative numbers $\omega_j$
such that
$\omega_j\ge\mathrm{Pr}(\bar v)\eta_j$
for every $j\in[m]$
and
$\sum_{j\in[m]}\omega_j\le\mathrm{Pr}(\bar v)$.
For each $j$, the preceding construction gives an upward flow on
$\{t^{j,k}:k\in[K_j]\}$ whose total flow into $\bar v$ is at most
$\omega_j$. Combining these flows over all $j$ preserves
\eqref{eq:flow:f1} and \eqref{eq:flow:f2} at every type other than
$\bar v$, while the last displayed inequality guarantees
\eqref{eq:flow:f1} at $\bar v$. Hence the combined flow is feasible,
top-preserving, and upward.

Thus a feasible top-preserving flow on
$\mathcal B\times\mathcal B$ exists if and only if there are
nonnegative numbers $\omega_j$ satisfying the two conditions above. 
Since $\mathrm{Pr}(\bar v)>0$, such numbers exist if and only if
\[
\sum_{j\in[m]}\eta_j\le1,
\]
which is precisely TTRB. Moreover, when TTRB holds, we may take
$\omega_j=\mathrm{Pr}(\bar v)\eta_j$
for every $j\in[m]$ 
and the construction above yields an upward feasible top-preserving
flow.
\end{proof}

\begin{proof}[Proof of Proposition~\ref{prop:regularity}]
Necessity follows from TTRB by taking $k=K_j-1$ for each item
$j\in[m]$. For sufficiency, use $N_{j,k}$ and $D_{j,k}$ as defined in
the proof of Proposition~\ref{prop:upward}. Fix $j\in[m]$.
If $v_{j,K_j-1}=0$, then $V_j=\{0,\bar v_j\}$ and $\rho_j=0$.
Otherwise,
$N_{j,K_j-1}=p_j(v_{j,K_j-1})v_{j,K_j-1}>0$, 
and
$D_{j,K_j-1}=p_j(\bar v_j)(\bar v_j-v_{j,K_j-1})$. 
Every summand in \eqref{eq:regular-terminal} is nonnegative, so that
condition implies $N_{j,K_j-1}\le D_{j,K_j-1}$.
Hence the factor \(1-D_{j,K_j-1}/N_{j,K_j-1}\) is nonpositive.
Myerson regularity gives
\[
\begin{aligned}
\phi_{j,\ell}
&\le\phi_{j,K_j-1}
=v_{j,K_j-1}\left(1-\frac{D_{j,K_j-1}}{N_{j,K_j-1}}\right)\\
&\le v_{j,\ell}\left(1-\frac{D_{j,K_j-1}}{N_{j,K_j-1}}\right)
\end{aligned}
\]
for every $\ell\in[K_j-1]$. 
Substituting the definition of
$\phi_{j,\ell}$ and rearranging gives
\[
D_{j,K_j-1}p_j(v_{j,\ell})v_{j,\ell}
\le N_{j,K_j-1}T_{j,\ell+1}(v_{j,\ell+1}-v_{j,\ell}).
\]
Summing from $\ell=k$ to $K_j-1$ yields
\[
\begin{aligned}
D_{j,K_j-1}N_{j,k}
&\le N_{j,K_j-1}\sum_{\ell=k}^{K_j-1}
T_{j,\ell+1}(v_{j,\ell+1}-v_{j,\ell})\\
&=N_{j,K_j-1}D_{j,k}
\end{aligned}
\]
for every $k\in[K_j-1]$. Dividing by the positive denominators shows
that $N_{j,k}/D_{j,k}$ is maximized at $k=K_j-1$. Thus
$\rho_j=N_{j,K_j-1}/D_{j,K_j-1}$. 
The same equality holds when \(v_{j,K_j-1}=0\), 
since in that case \(V_j=\{0,\bar v_j\}\) and both sides are zero. 
Summing this equality over $j\in[m]$, condition~\eqref{eq:regular-terminal}
gives the slice-tail condition~\eqref{eq:slice-tail}. Proposition~\ref{prop:upward}
provides a feasible upward flow, and Corollary~\ref{cor:slice-tail-sufficient}
gives optimality of top selling.
\end{proof}

\begin{proof}[Proof of Proposition~\ref{prop:boundary-dominance}]
Since $\bar v_j>0$ for every $j\in[m]$, divide each support value
$v_{j,k}$ by $\bar v_j$, keeping its probability unchanged.
The ratios $\eta_{j,k}$ and $v_{j,K_j-1}/\bar v_j$ and all relevant
probabilities are unchanged, so TTRB and boundary dominance are unchanged.
Under the corresponding bijection of types, top preservation and
\eqref{eq:flow:f1} are unchanged, while both sides of
\eqref{eq:flow:f2} for item $j$ are divided by $\bar v_j$.
Thus a feasible top-preserving flow exists before rescaling if and
only if one exists afterward. Proposition~\ref{prop:flow-characterization}
then shows that top-selling optimality is also preserved. We may
therefore assume $\bar v_j=1$ for every $j\in[m]$.
The construction in Proposition~\ref{prop:boundary-flow}
gives a feasible top-preserving flow $\lambda^B$ on
$\mathcal B\times\mathcal B$ satisfying
\eqref{eq:flow:f2} with equality for every $a\in\mathcal B$ and
$j\in[m]$ with $a_j<1$. For each $a\in\mathcal B$, define
\[
b(a):=\mathrm{Pr}(a)+\sum_{\hat t\in\mathcal B}
[\lambda^B(a,\hat t)-\lambda^B(\hat t,a)]\ge0.
\]
Thus $b(a)$ is the residual capacity for receiving flow at $a$ after
placing $\lambda^B$.
Nonnegativity follows from \eqref{eq:flow:f1} for $\lambda^B$.
For any event $A\subseteq\mathcal V$, write
$\mathrm{Pr}(A):=\sum_{t\in A}\mathrm{Pr}(t)$.
Let $A_h:=\{a\in\mathcal B:a_h<1\}$ for $h\in[m]$. These sets give the
disjoint partition
\[
\mathcal B=\{\bar v\}\cup A_1\cup\cdots\cup A_m.
\]
Since $m\ge2$ and every item has a value below its highest value with
positive probability, $\mathrm{Pr}(\mathcal V\setminus\mathcal B)>0$.
Summing the definition of $b$ over $\mathcal B$ gives
\[\sum_{a\in\mathcal B}b(a)=\mathrm{Pr}(\mathcal B).\]
For every $j\in[m]$,
\[
\begin{aligned}
\sum_{a\in\mathcal B}(1-a_j)b(a)
&=\sum_{a\in\mathcal B}(1-a_j)\mathrm{Pr}(a)+\sum_{a\in\mathcal B}\sum_{\hat t\in\mathcal B}
\lambda^B(a,\hat t)(\hat t_j-a_j)\\
&=\sum_{a\in A_j}[(1-a_j)+a_j]\mathrm{Pr}(a)\\
&=\mathrm{Pr}(A_j).
\end{aligned}
\]
The second equality uses top preservation when $a_j=1$ and equality
in \eqref{eq:flow:f2} when $a_j<1$.

For each $t\in\mathcal V\setminus\mathcal B$ and $a\in\mathcal B$, set
\[
\lambda^I(t,a):=
\frac{\mathrm{Pr}(t)}{\mathrm{Pr}(\mathcal V\setminus\mathcal B)}b(a)
\one_{\{a_h=1\ \forall h\in[m]:\,t_h=1\}},
\]
and set $\lambda^I(t,\hat t)=0$ for all other pairs $t,\hat t\in\mathcal V$.
The indicator restricts destinations to boundary types that preserve
every coordinate already at its top value in $t$.
Extend $\lambda^B$ by zero outside $\mathcal B\times\mathcal B$.
For each $a\in\mathcal B$,
\[
\sum_{t\in\mathcal V\setminus\mathcal B}\lambda^I(t,a)\le b(a),
\]
because the source probabilities sum to at most
$\mathrm{Pr}(\mathcal V\setminus\mathcal B)$. Thus \eqref{eq:flow:f1}
holds for $\lambda^B+\lambda^I$ at every $a\in\mathcal B$.
It also holds outside $\mathcal B$,
because $\lambda^B(\hat t,t)=\lambda^I(\hat t,t)=0$ for every
$t\in\mathcal V\setminus\mathcal B$ and $\hat t\in\mathcal V$.

Fix $t\in\mathcal V\setminus\mathcal B$ and a coordinate $j\in[m]$
with $t_j<1$. There is another coordinate $k\in[m]\setminus\{j\}$
with $t_k<1$, by the definition of $\mathcal B$. Define
\[
c:=\frac{\mathrm{Pr}(\bar v)}{p_k(\bar v_k)}
=\mathrm{Pr}(\bar v)+\mathrm{Pr}(A_k)\in(0,1).
\]
The equality holds because $\{\bar v\}\cup A_k$ is the event that all
items other than $k$ have their top values.
The strict bounds follow from $m\ge2$ and
$0<p_h(\bar v_h)<1$ for every $h\in[m]$. By
\eqref{eq:primitive-bd:pbd},
\[
t_j\le v_{j,K_j-1}
\le\frac{\mathrm{Pr}(\bar v)}{\max_{\ell\in[m]\setminus\{j\}}p_\ell(\bar v_\ell)}
\le c.
\]
The same condition gives
$a_h\le c$ whenever $a\in A_h$
and $t_h=1$, since such an $h$ differs from $k$. Consequently,
\[
(1-c)\sum_{a\in A_h}b(a)
\le\sum_{a\in A_h}(1-a_h)b(a)
=\mathrm{Pr}(A_h).
\]
The admissible destinations from $t$ are
$\mathcal B\setminus\bigcup_{h\in[m]:\,t_h=1}A_h$.
On $A_j$, use $a_j-t_j\ge(1-c)-(1-a_j)$. Outside $A_j$, use
$a_j-t_j=1-t_j\ge1-c$. The excluded sets $A_h$ are disjoint from
$A_j$ because $t_j<1$. Thus we subtract their contributions to
$(1-c)\sum_{a\in\mathcal B}b(a)$ and obtain
\[
\begin{aligned}
&\frac{\mathrm{Pr}(\mathcal V\setminus\mathcal B)}{\mathrm{Pr}(t)}
\sum_{a\in\mathcal B}\lambda^I(t,a)(a_j-t_j)\\
&\ge(1-c)\mathrm{Pr}(\mathcal B)-\mathrm{Pr}(A_j)
-(1-c)\sum_{\substack{h\in[m]\\t_h=1}}\sum_{a\in A_h}b(a)\\
&\ge(1-c)\mathrm{Pr}(\mathcal B)-\mathrm{Pr}(A_j)
-\sum_{\substack{h\in[m]\\t_h=1}}\mathrm{Pr}(A_h)\\
&=c\,\mathrm{Pr}(\mathcal V\setminus\mathcal B)
+\sum_{\substack{h\in[m]\setminus\{j,k\}\\t_h<1}}\mathrm{Pr}(A_h)\\
&\ge t_j\mathrm{Pr}(\mathcal V\setminus\mathcal B).
\end{aligned}
\]
The equality uses the displayed partition of $\mathcal B$ and
$\mathrm{Pr}(\bar v)+\mathrm{Pr}(A_k)=c$. 
The final inequality uses
$c\ge t_j$ and nonnegativity of the remaining probabilities.
Multiplying by
$\mathrm{Pr}(t)/\mathrm{Pr}(\mathcal V\setminus\mathcal B)$ proves the
inequality in \eqref{eq:flow:f2} for every
$t\in\mathcal V\setminus\mathcal B$ and $j\in[m]$ with $t_j<1$.
By construction, $\lambda^I(t,a)>0$ and $t_h=1$ imply $a_h=1$
for every $t\in\mathcal V\setminus\mathcal B$, $a\in\mathcal B$, and
$h\in[m]$, so $\lambda^I$ is top-preserving.
Since $\lambda^I(a,\hat t)=0$ for every $a\in\mathcal B$ and
$\hat t\in\mathcal V$, the left-hand side of \eqref{eq:flow:f2}
is unchanged there. The sum $\lambda^B+\lambda^I$ therefore gives a
feasible top-preserving flow
on the full type space. Proposition~\ref{prop:flow-characterization}
proves the claimed one-buyer optimality.
\end{proof}

\begin{example}[The sufficient conditions are incomparable]
\label{ex:sufficient-conditions-independent}
In both cases, $m=2$ and $F_1=F_2$.

First, let $V_j=\{1/1000,9/10,1\}$ for each $j\in[2]$, with
\[
\bigl(p_j(v_{j,1}),p_j(v_{j,2}),p_j(\bar v_j)\bigr)
=\left(\frac{49}{100},\frac1{100},\frac12\right).
\]
For either item,
\[
\begin{aligned}
\rho_{j,1}&=\frac{949}{50849}<\rho_{j,2}=\frac9{50},\\
\eta_{j,1}&=\frac1{999}<\eta_{j,2}=\frac9{50}.
\end{aligned}
\]
Thus the left-hand sides of \eqref{eq:slice-tail} and
\eqref{eq:ttrb} both equal \(9/25<1\), while boundary dominance
fails because \(v_{j,2}/\bar v_j=9/10>1/2=p_j(\bar v_j)\).

Next, let $V_j=\{3/5,2/3,1\}$ for each $j\in[2]$, with
\[
\bigl(p_j(v_{j,1}),p_j(v_{j,2}),p_j(\bar v_j)\bigr)
=\left(\frac{1-p}{4},\frac{3(1-p)}{4},p\right),
\]
where $3/4<p<25/33$. For either item,
\[
\begin{aligned}
\eta_{j,1}=\eta_{j,2}&=\frac{3(1-p)}{2p},\\
\rho_{j,1}&=\frac{13(1-p)}{1+7p}
>\rho_{j,2}=\frac{3(1-p)}{2p}.
\end{aligned}
\]
The left-hand side of \eqref{eq:ttrb} is \(3(1-p)/p<1\),
while the left-hand side of \eqref{eq:slice-tail} is
\(26(1-p)/(1+7p)>1\). Boundary dominance holds because
\(v_{j,2}/\bar v_j=2/3<p\). Hence TTRB and boundary dominance
together do not imply the slice-tail condition.
\hfill$\triangle$
\end{example}

\printbibliography
\end{document}